\documentclass[aps,pra,twocolumn,amsfonts,nofootinbib]{revtex4-1}
\pdfoutput=1
\usepackage[utf8]{inputenc}
\usepackage[dvips]{graphicx}
\usepackage{amsfonts}
\usepackage{amssymb}
\usepackage{amsmath} 
\usepackage{bm}
\usepackage{hyperref}
\usepackage{mathrsfs} 
\usepackage{float} 
\usepackage{slashed}
\usepackage{ulem}
 
\usepackage{mathdots}
\usepackage{braket}  
\usepackage{color} 
\usepackage{bbold}
\usepackage{anyfontsize}
\usepackage{lmodern}

\usepackage{capt-of}

\usepackage{makecell}
 
\usepackage{mathtools}

\newcounter{cquestion}

\newcommand{\nn}{\nonumber}

\renewcommand{\bar}[1]{\overline{#1}}
\renewcommand{\tilde}[1]{\widetilde{#1}}

\newcommand{\<}{\langle}
\renewcommand{\>}{\rangle}

\newcommand{ \pr }[1]{ \left( #1 \right) }

\renewcommand{\cal}{\mathcal}
\newcommand{\CO}{\mathcal{O}}

\newcommand{\CL}{\mathcal{L}}

\newcommand{\Ocal}{\mathcal{O}}
\newcommand{\fr}{\frac}
\newcommand{\p}{\partial}
\newcommand{\UV}{{\rm{UV}}}
\newcommand{\IR}{{\rm{IR}}}
\newcommand{\ra}{\rightarrow}
\newcommand{\TIM}{{\rm{TIM}}}
\newcommand{\Dmax}{\Delta_{\max}}
\newcommand{\de}{\delta}

\newcommand{\mkink}{m_{\textrm{kink}}}

\newcommand{\dmax}{\Delta_{\rm max}}

\newcommand{\mgap}{m_{\rm gap}}

\newenvironment{monospace}{\ttfamily}{\par}

\newcounter{cexample}[section]
\numberwithin{cexample}{section}

\newcommand{\bw}{\begin{widetext}}
\newcommand{\ew}{\end{widetext}}
\newcommand{\bea}{\begin{eqnarray}}
\newcommand{\eea}{\end{eqnarray}}
\newcommand{\be}{\begin{equation}}
\newcommand{\ee}{\end{equation}}

\numberwithin{equation}{section}
\begin{document}

\title{Solving the 2D SUSY Gross-Neveu-Yukawa Model \\ with  Conformal Truncation}

\author{A. Liam Fitzpatrick$^1$, Emanuel Katz$^1$, Matthew T. Walters$^{2,3}$, Yuan Xin$^1$}
\affiliation{$^1$Boston University, Boston, Massachusetts 02215, USA \\
$^2$Theoretical Physics Department, CERN, 1211 Geneva 23, Switzerland \\
$^3$Institute of Physics, \'Ecole Polytechnique F\'ed\'erale de Lausanne (EPFL), CH-1015 Lausanne, Switzerland}

\date{\today}

\begin{abstract}
We use Lightcone Conformal Truncation to analyze the RG flow of the two-dimensional supersymmetric Gross-Neveu-Yukawa theory, i.e.\ the theory of a real scalar superfield with a $\mathbb{Z}_2$-symmetric cubic superpotential.  The theory depends on a single dimensionless coupling $\bar{g}$, and is expected to have a critical point at a tuned value $\bar{g}_*$ where it flows in the IR to the Tricritical Ising Model (TIM);  the theory spontaneously breaks the $\mathbb{Z}_2$ symmetry on one side of this phase transition, and breaks SUSY on the other side.   We calculate the spectrum of energies as a function of $\bar{g}$ and see the gap close as the critical point is approached, and numerically read off the  critical exponent $\nu$ in TIM.  Beyond the critical point, the gap remains nearly zero, in agreement with the expectation of a massless Goldstino.  We also study spectral functions of local operators on both sides of the phase transition and compare to analytic predictions where possible.  In particular, we use the Zamolodchikov $C$-function to map the entire phase diagram of the theory.   Crucial to this analysis is the fact that our truncation is able to preserve supersymmetry sufficiently to avoid any additional fine tuning.

 \end{abstract}

\maketitle
\tableofcontents

\section{Introduction}
 Quantum Field Theory (QFT) has an astonishingly broad range of applicability, yet is notoriously difficult at strong coupling. Hamiltonian truncation methods are a promising approach for computing real-time dynamical quantities in generic QFTs, but much work remains to be done to develop them more fully.  In this paper, we will work with a specific Hamiltonian truncation method, Lightcone Conformal Truncation (LCT)~\cite{Katz:2013qua,Katz:2014uoa,Katz:2016hxp,Phi4Paper,Delacretaz:2018xbn}, which has a number of advantages but also comes with additional challenges.  We will focus on a specific model, the 2D Supersymmetric Gross-Neveu-Yukawa (SGNY) model, 
as a useful case to explore issues that arise when studying theories with both scalars and fermions. The 2D SGNY model is a theory of a real superfield 
\be
\Phi = \phi + \bar{\theta} \psi + \theta \chi + \theta \bar{\theta} F,
\ee 
and a  superpotential
\be
W(\Phi) = h \Phi + \frac{g}{3!} \Phi^3.
\label{eq:FirstSPotl}
\ee
The scalar potential is $\frac{1}{2} (W'(\phi))^2$, and the scalar-fermion coupling is $W''(\phi) \psi \chi$.  
 Much is known already about this theory, which will permit many nontrivial checks of our numerical results.

 LCT is a numeric method for studying QFT nonperturbatively.  The basic idea is to numerically diagonalize the  lightcone Hamiltonian $P_+$, i.e.\ the generator of translations in the lightcone direction $x^+ = (x^0+ x^1)/\sqrt{2}$, in a basis of operators that, roughly speaking, have scaling dimensions below some truncation limit $\Delta_{\rm max}$ in the ultraviolet (UV). The UV is taken to be a known, solvable CFT, and the full theory is the UV CFT deformed by one or more relevant operators:
 \be
 P_+ = P_+^{(\rm CFT)} + \sum_i g_i \int d^{d-1} x \ \CO_i(x).
 \label{eq:GenPP}
 \ee
Many interesting models, including the SGNY model, are of this form.  The UV CFT of SGNY is just  a free massless scalar and fermion.  For a pedagogical introduction to the setup and methods of LCT, we refer the reader to the companion paper \cite{pedagogical}.   

One of the main innovations that we will employ in this paper is to use a modified definition of the truncated Hamiltonian which uses the SUSY algebra.  In $d=2$, ${\cal N}=(1,1)$ SUSY, there are two supercharges $Q_\pm$, and they satisfy
\be
Q_\pm^2 =  P_\pm ,
\ee
where we have chosen a specific convention for the normalization of $Q_\pm$.  Rather than computing the matrix elements of $P_+$, we can compute the matrix elements of $Q_+$ in our truncated basis and then square it.\footnote{This same approach has been used to study supersymmetric theories in the context of Discrete Light Cone Quantization~\cite{Matsumura:1995kw,Antonuccio:1998kz,Antonuccio:1998jg,Antonuccio:1998tm,Antonuccio:1998mq,Antonuccio:1998zp,Antonuccio:1998xs,Antonuccio:1998zu,Antonuccio:1998mw,Antonuccio:1999iz,Lunin:1999ib,Hiller:2000nf,Pinsky:2000rn,Lunin:2000im,Filippov:2000iu,Hiller:2001qb,Hiller:2002pj,Hiller:2003qe,Harada:2003bs,Harada:2004ck,Hiller:2004vy,Harada:2004wu,Hiller:2004rb,Hiller:2005vf,Hiller:2007sc,Trittmann:2009dw}, where the $x^-$ direction is compactified. That method has largely focused on gauge theories and, to the best of our knowledge, has not been used to study the 2D SGNY model.} One advantage of this approach is that $Q_+$ is much simpler than $P_+$.  In general,
\be
 Q_+ \propto \int d x^- W'(\phi) \psi.
 \label{eq:QpGen}
\ee
Thus, $Q_+$ contains fewer terms than $P_+$, and on the LC it is also local (unlike the Hamiltonian).
However, the main advantage of using the supercharge is the ability to preserve SUSY sufficiently in LCT,  in a manner which avoids having to fine-tune UV-divergent counterterms to maintain the symmetry.  Indeed, in a naive use of Hamiltonian methods, one normal-orders, generically leading to a breaking of SUSY.

Our main results include a numerical computation of the mass spectrum as a function of the dimensionless coupling 
\be
\bar{g} \equiv \frac{g}{m}
\ee 
in (\ref{eq:OurSPotl}), as well as of spectral functions $\rho_{\cal O}(\mu^2)$ of various operators $\CO$ as a function of invariant mass-squared $\mu^2$.  We will pay special attention to the spectral function of the stress-tensor, $T$, since its integral is Zamolodchikov's $C$-function~\cite{Zamolodchikov:1986gt,Cappelli:1990yc}.  As this function appears in various thermodynamic quantities characterizing the QFT, we may regard its dependence on $\mu$ as being equivalent to its dependence on temperature.  Hence, a map of the $C$-function as a function of $\mu$ and $\bar{g}$ can be considered a representation of the phase diagram of the theory.   This map is shown in Fig.\ \ref{fig:heatPlot} which captures the more qualitative phase diagram illustrated in Fig.\ \ref{fig:PhaseDiagramCartoon}.

The coupling dependence of the mass spectrum is shown in  Fig.\ \ref{fig:spectrum}. We discuss its interpretation in Section \ref{sec:mass-spectrum-interacting}.  From the spectrum we read off the critical coupling $\bar g_*$ where the SGNY theory flows to the TIM critical point. In the $\mathbb{Z}_2$-breaking phase, we fit the closing of the mass gap near the critical point as a function of $\bar{g}$ to extract the critical exponent $\nu$. Fig.\ \ref{fig:criticalExponent} displays the convergence of the critical exponent $\nu$ to the theoretical expectation $\nu=1.25$ as $\dmax$ increases.  The Zamolodchikov $C$-function and the trace of $T$ near the critical point are shown in Fig.\ \ref{fig:CFuncGrid}
and Fig.\ \ref{fig:TraceGrid} respectively.  Included in these plots are the theoretical curves obtained from TIM integrability results for comparison.
The numerical spectrum Fig.\ \ref{fig:spectrum} also shows that the SGNY theory flows to the massless SUSY-breaking phase at large $\bar g$. In the IR the numerical $C$-function approaches the central charge $c_{\rm Ising} = 0.5$ of the IR fixed point, shown in Fig.\ \ref{fig:ising-extrapolation}.  We also check that other correlators agree with Ising CFT behavior in Fig.\ \ref{fig:ising-phiAndPhi2}.

The outline of the paper is as follows. In Section \ref{sec:rg-flow} we review some key features of the RG flows of the SGNY theory. We will discuss how the SGNY model is defined in the UV, what phases we are expecting in the IR, and the properties of the critical point. In Section \ref{sec:setup} we set up the Hamiltonian truncation framework. We will compute the Hamiltonian matrix elements of the relevant deformation with respect to the conformal basis in lightcone quantization. We warm up by discussing the SGNY theory and Hamiltonian truncation in the free and perturbative regimes. 
We present the numerical results of the strongly coupled SGNY theory in Section \ref{sec:massive} and \ref{sec:susy-breaking-phase}. 
Section \ref{sec:massive} focuses on the gapped $\mathbb{Z}_2$-breaking phase.
In the first subsection \ref{sec:mass-spectrum-interacting} we display the mass spectrum, read off the phase structure from the spectrum and discuss the convergence of the numerics. 
In the following subsections \ref{sec:criticalExponent}, \ref{sec:CFunc} and \ref{sec:Trace} we zoom-in to the vicinity of the TIM critical point, and compute the critical exponent, the Zamolodchikov $C$-function and the spectral function of the trace of the stress tensor, respectively. 
In the subsection \ref{sec:universal-IR} we show that the truncation effects have universal behavior in the IR. 
In Section \ref{sec:susy-breaking-phase} we move on to the SUSY-breaking phase, which has an IR fixed point in the universality class of the 2D Ising model. We provide evidence that the spectrum and the spectral functions of various operators match the Ising CFT in the IR.

\section{The RG Flow and Infrared}
\label{sec:rg-flow}

\begin{figure}[h!]
\begin{center}
\hspace{-1cm}
\includegraphics[width=0.85\columnwidth]{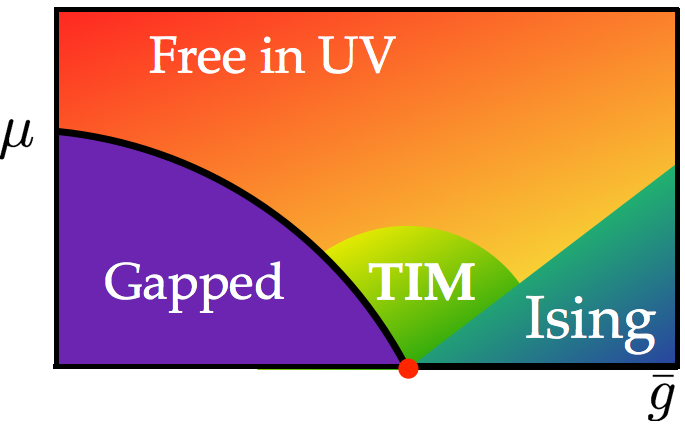}
\caption{Cartoon of the phase diagram of SGNY, as a function of the dimensionless parameter $\bar{g}$ and energy scale $\mu$. The thick solid line indicates the gap, while the color gradients indicate RG flows between different regimes.  A more precise version of this diagram obtained from our numerical results, with a similar color scheme, is exhibited in Fig.\ \ref{fig:heatPlot}.}
\label{fig:PhaseDiagramCartoon}
\end{center}
\end{figure}

Before we describe the truncation setup in more detail, here we will review some key features of the SGNY theory; for more details, see \cite{Kastor:1988ef,Friedan:1984rv}.  By inspection of the superpotential (\ref{eq:FirstSPotl}), the theory has an interesting phase structure depending on the value of the dimensionless ratio $g/h$.  In addition to supersymmetry, the Lagrangian has a chiral $\mathbb{Z}_2$ symmetry under which $\phi$ and $\psi$ are odd but $\chi$ is even. For large positive $h/g$, the vacuum has $\< \phi\>=0, \< W'(\phi)\> \ne 0$ so SUSY is broken spontaneously while the chiral $\mathbb{Z}_2$ is not, whereas at large negative $h/g$, the vacuum has $\< \phi\> \ne 0, \< W'(\phi)\> =0$ so the reverse is true. On the spontaneous SUSY breaking side, the theory has a massless Goldstino, and so flows to the 2D Ising model in the infrared (IR).  At the transition between these two phases lies the Tricritical Ising Model (TIM).  A cartoon depicting this expectation is shown in Fig.\ \ref{fig:PhaseDiagramCartoon}.

 The critical point of TIM is the unique CFT that shows up in both the nonsupersymmetric and ${\cal N}=1$ minimal series of 2D CFTs.  When the interaction $W(\phi) \supset g \Phi^3$ is turned on,  $\Phi^2$ is a descendant of $\Phi$ by the equations of motion and the only relevant primary operators in the weakly coupled regime are $1$ and $\Phi$, as well as spin operators $\sigma$ and $\sigma' \sim \Phi \sigma$ defined as boundary-condition-changing operators for the fermions.  Because $\sigma, \sigma'$ cannot be constructed from local products of the components of $\Phi$ acting on the vacuum, they do not appear in our construction and we will not see them at any point along the RG flow.    In the IR, $\Phi$ flows to the $\Delta= \frac{1}{5}$ operator $\epsilon$ in TIM.  TIM also has scalar operators $\epsilon'$ and $\epsilon''$ with $\Delta=\frac{6}{5}$ and $\Delta=3$ that are primaries under the Virasoro algebra, but are descendants of $\epsilon$ and 1, respectively, under the super-Virasoro algebra.  

\begin{center}
\begin{table}[h!]
\renewcommand{\arraystretch}{1.5}
\begin{tabular}{|c|c ||c|c|}
\hline
$\Delta_{\rm UV}$ & $\CO_{\rm UV}$ & $\Delta_{\rm IR}$ & $\CO_{\rm IR}$ \\
\hline
$0$ &  $\phi$ &  $\frac{1}{5}$ & $\epsilon$\\
\hline
$1$ &  $W'(\phi)$ &  $\frac{6}{5}$ & $\epsilon'$\\
\hline
$3$ & $(\partial_\mu \phi)(\partial^\mu \phi) \psi \chi$ & 3 & $\epsilon''$ \\
\hline
\end{tabular}
\label{tab:TIMContent}
\caption{UV to IR matching of scalar Virasoro-primary operators in the flow from the free theory to the critical point of TIM. 
}
\end{table}
\end{center}
The flow from the critical point of TIM to Ising is triggered by deforming by $\epsilon'$, so supersymmetry is broken spontaneously.  This deformation is described in the IR as the term $h \Phi$ in the superpotential (\ref{eq:FirstSPotl}).   If the sign of the coefficient is flipped, the theory flows to a massive phase with preserved SUSY. By scaling, the $\epsilon'$ deformation to the massive phase can be written to leading order in the deformation  in two equivalent ways:
\be
\delta \CL \propto (g_*-g) \left( \frac{1}{2} \phi^2 + \psi \chi\right) \propto m_{\rm gap}^{2-\Delta_{\epsilon'}} \epsilon',
\ee
 where $g_*$ is the critical coupling.  Therefore, the gap closes as a function of $g$ near the critical point as
\be\label{eq:criticalExponent}
m_{\rm gap} \propto (g_*-g)^{\nu}, \qquad \nu = 1.25.
\ee

In fact, the $\epsilon'$ deformation around TIM is integrable, and it is in principle possible to compute correlators along the resulting RG flow nonperturbatively.  On the gapped side, the massive particles can be thought of as `kink' states created by a profile in $\phi$ that interpolates between two minima.  We will use such integrability results taken from \cite{Delfino:1999et} to compare to our numeric results for correlators of the theory not only at the critical point but also in the neighboring region on either side of it.

\section{Conformal Truncation Setup}
\label{sec:setup}

In this section, we will describe how we construct the conformal truncation Hamiltonian in lightcone quantization.  We will also work through a few perturbative computations explicitly.  These perturbative computations will allow us to provide some intuition analytically, and also to perform a few consistency checks, before moving on to completely numeric results in the subsequent sections.

\subsection{Lightning Review of Conformal Truncation}

In LCT, we label our basis of states by their momentum and the primary operator whose irrep they appear in under the conformal algebra.  The Hamiltonian $P_+$ in lightcone quantization does not mix states of different spatial momentum $P_-$ and thus we always work at a fixed value of $P_-$ for all physical states, which by boosting we may set to $P_-=1$ without loss of generality. As mentioned in the introduction, we consider theories that are deformations of a UV CFT by one or more relevant operators, and we use the primary operators of the UV CFT for the basis.  In this paper, the UV CFT is a theory of a free massless superfield, i.e.\ a massless real scalar $\phi$ and real fermion $(\psi,\chi)$. The primary operators are therefore built from the current $\p_\pm\phi$, vertex operators $V_\alpha \equiv e^{ \alpha \phi}$, and the fermion components $\psi,\chi$. In the free theory, the equations of motion set $\partial_+ \partial_- \phi=0$, so in momentum space $p_+ p_-=0$ for the scalars. Similarly, the equations of motion for fermions set $p_+ = 0$ for $\psi$ and $p_-=0$ for $\chi$. In LC quantization, we integrate out the $p_-=0$ ``zero modes'', which are nondynamical owing to the fact that they have no time derivative in their kinetic term, removing operators built from $\p_+\phi$ and $\chi$.

There is a further complication, however: because the relevant deformations we consider have $\Delta \leq \fr{d}{2}$, there are IR divergences in the resulting Hamiltonian matrix elements. The effect of these divergences is to remove all vertex operators from the basis, as well as any operators without derivatives acting on $\psi$. Conveniently, the free fermionic theory with all factors of $\psi$ projected out is a Generalized Free Theory (GFT) where $\partial_- \psi$ is a primary operator with $h=3/2$.  Nevertheless, even for free theories and GFTs, explicitly constructing all the primary operators up to large dimension is a nontrivial task; the methods we employ for constructing this basis, as well as the details of the removal of states due to IR divergences, are described in \cite{pedagogical}. 


In summary, the basis consists of operators $\CO_i$ constructed from all products of $\partial_- \phi$ and $\partial_- \psi$. The states themselves are constructed by Fourier transforming these operators acting on the vacuum, 
\be
| \CO_i\> \equiv \int d^2 x \, e^{-i P \cdot x} \CO_i(x) | \textrm{vac} \>,
\ee
so that they have definite momentum. 
It follows that the matrix elements of a deformation (\ref{eq:GenPP}) are integrals over three-point functions~\cite{Anand:2019lkt}:
\begin{equation}
\begin{aligned}
&\< \CO_i | \delta P_+ | \CO_j\> \\
&= \sum_k g_k \! \int \! d^2x \, dy^- \, d^2 z \, e^{i(P\cdot x - P'\cdot z)}  \< \CO_i(x) \CO_k(y) \CO_j(z)\>.
\end{aligned}
\end{equation}

Once we diagonalize $P_+$, we can construct spectral functions of local operators $\CO$, as follows.  Let $|\mu_k\>$ be the eigenvectors of $P_+$:
\be 
2 P_- P_+ |\mu_k\> = \mu_k^2 | \mu_k\>.
\ee
The spectral function $\rho_\CO$ of an operator $\CO$ is
\be
\rho_\CO(\mu^2) =  \sum_k | \< \textrm{vac} | \CO(0) | \mu_k\>|^2 \delta(\mu^2-\mu_k^2) .
\label{eq:GenSpecFunc}
\ee
By diagonalizing $P_+$,  we obtain the overlaps in the above formula for each eigenvector $|\mu_k\>$ and local operator $\CO$.

\subsection{Constructing the Supersymmetric Lightcone Hamiltonian}

In this subsection, we describe the construction of the LC Hamiltonian and the computation of its matrix elements.  For reasons we discuss below, instead of using the form of the superpotential (\ref{eq:FirstSPotl}), we will perform a field redefinition  $\Phi \rightarrow \Phi+c$ to absorb the linear term $h \Phi$.  We can parameterize the resulting superpotential as
\be
W(\Phi) = \frac{1}{2} m \Phi^2 + \frac{g}{3!} \Phi^3.
\label{eq:OurSPotl}
\ee
When $h>0$ -- i.e.\ the SUSY-breaking, $\mathbb{Z}_2$ respecting phase --  the shift in $\phi$ to remove the linear term is imaginary and  it is not obvious that this new form of the superpotential is equivalent to the original one.  However, we will find numeric evidence that the form (\ref{eq:OurSPotl}) does indeed correctly produce the physics of the SUSY-breaking phase.  This is perhaps not too surprising since a model only needs to be able to dial the coefficients of all relevant operators allowed by symmetry in order to be in the right universality class.

\subsubsection{Standard Construction}

Our first task in constructing the LC Hamiltonian is to integrate out the component $\chi$ of the fermion, which has no time derivatives in the Lagrangian and is therefore nondynamical.  Before integrating out $\chi$, the Lagrangian is
\begin{eqnarray}
\CL &=& \frac{1}{2}(\partial \phi)^2 + \sqrt{2} i (\psi \partial_+ \psi + \chi \partial_- \chi) \nn\\
&+&  2i(m+ g \phi) \psi \chi  - \frac{1}{2} (m \phi + \frac{g}{2} \phi^2)^2 .
\end{eqnarray}
 We integrate out $\chi$ by solving its equations of motion, with the resulting nonlocal action:
\begin{eqnarray} 
\label{eq:PplusLag}
\CL &=& \frac{1}{2}(\partial \phi)^2 + \sqrt{2} i \psi \partial_+ \psi \\
&+&  \frac{i}{\sqrt{2}}(m \psi + g \phi \psi) \frac{1}{\partial_-} (m\psi+ g \phi \psi)  - \frac{1}{2} (m \phi + \frac{g}{2} \phi^2)^2 . \nn
\end{eqnarray}
The deformed Hamiltonian $\delta P_+$ has a total of 6 terms coming from the second line: two mass terms, two cubic terms, and two quartic terms.  SUSY is preserved up to truncation effects when the coefficients of these terms are set according to the above Lagrangian.  Although it is possible to use this form of the theory for LC truncation, in the next section we will turn to another construction that automatically implements SUSY and has a number of practical advantages.

 \subsubsection{Construction Through \texorpdfstring{$Q_+$}{Q+}}
 
 Because the 2D SUSY algebra equates the LC Hamiltonian $P_+$ and the square $Q_+^2$ of the supercharge, we can also obtain a truncated form of the Hamiltonian by first computing $Q_+$.  $Q_+$ is given by (\ref{eq:QpGen}), which in our case is
 \be\label{eq:QPlusMatrix}
 Q_+ = \sqrt{2} \int dx^- (m \phi + \frac{g}{2} \phi^2) \psi.
 \ee
By contrast with the construction in the previous subsection, there are only two terms here.  This simplification saves a significant amount of effort and computation time, but more importantly it leads to a number of qualitative simplifications as we will see.

Once we have computed the matrix elements of $Q_+$ in our truncation basis, we can {\it define} our Hamiltonian through the algebra:
\be\label{eq:Hamiltonian}
\< \CO_i | P_+ | \CO_j \> \equiv \sum_k \< \CO_i | Q_+  | \CO_k \> \< \CO_k | Q_+ | \CO_j\>.
\ee
Note that this definition is a modification of $P_+$, because the sum on $k$ is only over states in the truncation rather than over all states in the space.  It is perhaps useful to imagine taking two different truncation spaces, one $\Delta_{\rm max}$ for the external states $| \CO_i\>, | \CO_j\>$ in the above equation, and a separate $\Delta_{\rm max}^{\rm int}$ for the ``internal'' states in the sum over $k$.  In the limit that $\Delta_{\rm max}^{\rm int} \rightarrow \infty$, we would reproduce our previous definition of $P_+$. In practice, we will always use the same truncation space for both internal and external states. As we discuss in detail  in Section \ref{sec:perturbation}, UV divergences to the mass term are removed when we use $Q_+^2$ to define $P_+$. Such divergences in the original $P_+$ construction turn out to be a significant source of difficulty for studying the supersymmetric critical point, and their absence in the $Q_+^2$ construction is therefore almost crucial to our analysis. One notable aspect of the $Q_+^2$ construction is that $P_+$ is no longer simply the matrix elements of the exact Hamiltonian restricted to a subspace, and so our truncation is not strictly speaking a variational method approximation.  Consequently, the smallest eigenvalue of our new truncation $P_+$ can in principle be below the true smallest eigenvalue.  

 We will also make use of the generator $Q_-$, which takes the form
 \be
 Q_- = 2 \int dx^- (\partial_- \phi) \psi
 \ee
 independently of the interaction terms. The SUSY algebra imposes $Q_-^2 = P_-$.  Because we work in a frame where all states have $P_-=1$, we therefore have the useful fact that $Q_-$ squares to the identity.  The relation $Q_-^2 =1$ does not hold exactly because of truncation effects: $Q_-$ sometimes acts on states to raise their dimension and thereby takes states within the truncation space to states outside of it.  However, we can mitigate these effects somewhat by modifying our truncation.  In particular, note that $Q_-$ does not change particle number, so the truncation effects that violate $Q_-^2 \sim 1$ will be less severe if we choose a truncation that counts the number of $\phi$s and $\psi$s equally.  So, we define a modified maximum dimension $\Delta_{\rm max}$ for each operator that treats each  $\psi$ as if it had dimension 0, i.e.\ the same as $\phi$:
 \be
 \tilde{\Delta}[\partial^{k_1} \phi \dots \partial^{k_n} \phi \partial^{k_{n+1}} \psi \dots \partial^{k_m} \psi] \equiv \sum_{i=1}^m k_i .
 \ee
In other words, $\tilde{\Delta}$ is just the total number of derivatives in the operator. For all numeric results, we will impose $\tilde{\Delta} \le \Delta_{\rm max}$ as our truncation on the operators.

\subsection{Weak Coupling Warm-up}

Having set up the truncated Lightcone Hamiltonian for the SGNY model, we can now diagonalize it and start making physical observations. The simplest observables are the eigenvalues of $P_+$, i.e.\ the mass spectrum of particles and bound states.  We will also use the eigenvectors of $P_+$ to extract spectral functions of real-time correlators, per (\ref{eq:GenSpecFunc}). In this section, we will first warm up with free theory and perturbation theory, before turning to strong coupling in later sections.  The perturbative warm-up will also have the advantage of giving us analytic insight into the UV divergences of the theory; because the theory is super-renormalizable, all such divergences can be seen explicitly at low order in perturbation theory.

\subsubsection{Free Theory}
In lightcone quantization, the free theory Hamiltonian conserves particle number, so we can analyze each particle number sector separately.\footnote{This property is not shared by equal-time quantization, where mass terms $\phi^2$ and $\psi \chi$ can change particle number by $0$ or $\pm 2$.}  The states $|\partial \phi\>$ and $|\partial \psi\>$ are the only one-$\phi$ and one-$\psi$ states, respectively, so diagonalizing the Hamiltonian $P_+$ in the one-particle sector is trivial. The free theory $Q_+$ is 
\be
Q_+ = \sqrt{2} \int dx^- m \phi \psi,
\ee
and consequently $\frac{\< \partial \phi | Q_+ | \partial \psi\>}{\sqrt{\< \partial \phi| \partial \phi\> \< \partial \psi| \partial \psi\>}} =  m/\sqrt{2}$  in the free theory, so $Q_+^2 = P_+ = m^2/2$ on the one-particle sector as required by the SUSY algebra.

In the two-particle sector, we can have either two $\phi$s, two $\psi$s, or one $\phi$ and one $\psi$. In the first [second] case, there is one operator at each even [odd] integer degree\footnote{By `degree', we mean the number $\tilde{\Delta}$ of total derivatives in the operator minus the particle number $n$.  So e.g.\ the operator $\partial \phi \partial \psi$ has degree $k=0$.} $k\ge 0$, whereas in the third case there is one at every integer $k\ge 0$. We can therefore uniquely label each two-particle state by its degree and particle content.  For instance, the two-particle states up to degree $k \le 1$ are
\bea
\left[\phi \phi\right]_0 &\propto& \partial \phi \partial \phi, \\
\left[\phi \psi\right]_0 & \propto& \partial \phi \partial \psi, \\
\left[\phi \psi\right]_1 & \propto& 2\partial \phi \partial^2 \psi-3 \partial^2 \phi \partial \psi, \\
\left[\psi \psi\right]_1  &\propto& \partial \psi \partial^2 \psi.
\eea
The generator $Q_-$ takes $\phi \rightarrow \psi \rightarrow \partial \phi$. By inspection, it takes $[\phi \phi]_0 \rightarrow [\phi \psi]_0$, and due to (anti)symmetry of (fermions) bosons, it takes $[\phi \psi]_0$ into a total $\partial_-$ derivative of $[\phi \phi]_0$.  In momentum space, $\partial_-$ is just a constant, so $Q_-$ takes $|[\phi \phi]_0 \> \leftrightarrow |[\phi \psi]_0\>$ acting on our lightcone basis states.  By contrast, acting on two-particle states at higher degree $k>0$, $Q_-$ mixes states of different degree; e.g.\ it takes $[\psi \psi]_1$ to a linear combination of $[\phi \psi]_1$ and $[\phi \psi]_2$. More generally,  $Q_-$ can act on states to increase their degree, and therefore their modified dimension $\tilde{\Delta}$, by at most 1.  For states with $\tilde{\Delta}= \Delta_{\rm max}$ at the upper limit of the truncation, the action of $Q_-$ may\footnote{In some special cases, $Q_-$ keeps subsectors within the truncation space.  For instance, $Q_-$ 
mixes two-particle states at degree $2n-1$ and $2n$ for integer $n$, so in the two-particle subsector $Q_-$ is preserved by the truncation if $\Delta_{\rm max}$ is even and broken by the truncation if $\Delta_{\rm max}$ is odd.} take them out of the truncation subspace.  Therefore our truncation explicitly breaks SUSY.  Fortunately,  in LC quantization the only UV divergences in the theory  are logarithmic divergences; power-law divergences of the vacuum energy and tadpoles that would be present in  equal-time quantization require particle production from the vacuum, which is not possible in LC quantization. Logarithmic divergences receive only a $1/\Delta_{\rm max}$ suppressed contribution  from the last layer of modes near the truncation (since $\delta \log \Delta/\delta \Delta \sim \Delta^{-1}$), where some states are missing their $Q_-$ superpartners,   so this breaking is fairly mild.

Next, we illustrate a simple explicit example where we can see how $Q_+^2$ approximates $P_+$  at finite truncation.  The matrix element of $2P_+$ on the two-$\phi$ state $[\phi \phi]_0$ is
\be
\< [ \phi \phi ]_0 | 2 P_+ | [\phi \phi]_0 \> = 6 m^2,
\ee
where implicitly we have divided out the normalization $(2\pi) 2p_- \delta(p_--p_-')$ of the external states.
Mixing with higher degree two-$\phi$ states lowers the mass-squared of the lightest two-$\phi$ state to $4 m^2$, as one can see from the formulas in appendix \ref{app:FreeAndPert}. 
 For now, we mainly want to see explicitly in a simple example that as the truncation is lifted, the individual matrix elements of $Q_+^2$ approaches those of $P_+$.  $Q_+$ only mixes $|[ \phi \phi]_0\>$ with the $| [ \phi \psi]_k\>$ states, and the matrix elements are
\be
\< [\phi \phi]_0 | Q_+ | [\phi \psi]_k\> = m \sqrt{\frac{12}{(1+k)(2+k)(3+k)}}.
\ee
Consequently,
\be
\sum_{k\le K } |\< [\phi \phi]_0 | Q_+ | [\phi \psi]_k\>|^2 = 3m^2 \left( 1- \frac{2}{(2+K)(3+K)} \right),
\ee
which does indeed approach $\< [ \phi \phi ]_0 | P_+ | [\phi \phi]_0 \>$ at $K\rightarrow \infty$.

For a generic state $\ket{\Psi}$, the supersymmetry algebra promotes it to a supermultiplet
\be
Q_-^{s_-} Q_+^{s_+} |\Psi\>, \quad s_\pm = 0,1.
\ee
which generally gives a 4-fold degenerate mass eigenvalue, if all four states are linearly independent. When a state is annihilated by a linear combination of $Q_+$ and $Q_-$, i.e.\ the state is a BPS state, it will be 2-fold degenerate.  The truncation effects mentioned above break the two-fold degeneracy associated with $Q_-$ so that it is only approximate at finite $\Delta_{\rm max}$.  In contrast, the two-fold degeneracy associated with $Q_+$ is exact, since for any eigenvector $|\Psi\>$ of $P_+ = Q_+^2$, the state $Q_+ |\Psi\>$ will always be another eigenvector with the same eigenvalue.  In some special circumstances -- e.g.\ the two-particle sector in the free theory with even $\dmax$ -- both $Q_+$ and $Q_-$ are preserved exactly, and then most states have an exact 4-fold degeneracy.  
 As an example, Fig.\ \ref{tab:free-spectrum} shows the result for $\Delta_{\rm max}=8$ at 2- and 3-particle sectors in the free theory. The mass eigenvalues form a discrete sample of the  $n$-particle continuum.  Interestingly, in this case, in the 2-particle threshold there is a BPS state with only a 2-fold degeneracy, with mass eigenvalue exactly $4 m^2$.

\begin{figure}[t]
	\centering
	$$
	\begin{array}[t]{ll}
	\textbf{2 particles}& \\
	\hline
	2Q_+^2& 2P_+\\
	\hline
	 4.00000 & 4.13928 \\
	 4.00000 & 4.13928 \\
	 4.60752 & 4.60752 \\
	 4.60752 & 4.60752 \\
	 4.60752 & 5.52627 \\
	 4.60752 & 5.52627 \\
	 7.38800 & 7.38800 \\
	 7.38800 & 7.38800 \\
	 7.38800 & 10.9492 \\
	 7.38800 & 10.9492 \\
	 21.0045 & 21.0045 \\
	 21.0045 & 21.0045 \\
	 21.0045 & 51.3852 \\
	 21.0045 & 51.3852 \\
	\hline
	\end{array}
	\hspace{20px}
	\begin{array}[t]{ll}
	\textbf{3 particles}& \\
	\hline
	2Q_+^2&2P_+\\
	\hline
	 9.26887 & 9.80095 \\
	 9.26887 & 9.85907 \\
	 11.0727 & 11.1472 \\
	 11.0727 & 11.2936 \\
	 11.5068 & 12.9964 \\
	 11.5068 & 13.8108 \\
	 11.6121 & 14.4995 \\
	 11.6121 & 14.7542 \\
	 15.8602 & 16.2540 \\
	 \text{...} & \text{...} \\
	 32.2736 & 55.2074 \\
	 32.2736 & 57.3414 \\
	 38.6102 & 61.1966 \\
	 38.6102 & 68.6745 \\
	\hline
	\end{array}
	$$
	\caption{Mass-squared eigenvalues of the free theory in 2- and 3-particle sector, respectively, from diagonalizing $Q_+^2$ and $P_+$ in the truncated basis up to maximum degree $\dmax=8$. The truncated basis has 14 and 28 states,
	respectively, in each sector.
	The numbers in the table are in units $m^2$, where $m$ is the mass of a single particle.}
	\label{tab:free-spectrum}
\end{figure}

Finally, we end our discussion of the free theory with comments about the effect of the truncation on the spectrum of multi-particle states.
  As we show in appendix \ref{app:FreeAndPert}, the spectrum of two-$\psi$ states at large $\dmax$ is approximately 
\be
m^2_n = 4 m^2 \sec^2\left( \frac{2 \pi n}{2 \dmax + 7 } \right), \quad 0\le n \le \dmax/2.
\label{eq:freespectrumpsi}
\ee
From this expression, we see that the truncation has both UV and IR effects.  The UV effect is that the spectrum of two-$\psi$ states only goes up to $m_{\rm max}^2 \sim m^2 \dmax^2$, so $\dmax$ behaves like a UV cutoff as expected.  The IR effect is that the free theory spectrum of two-$\psi$ states near its threshold $4m^2$ is discretized approximately as $m_n^2 -4m^2 \sim m^2 \frac{n^2}{\dmax^2}$.  Roughly speaking, we can think of this IR truncation effect as putting the system in a box of size $\dmax^{-2}$.  Once we go to strong coupling, we will see additional IR truncation effects.

 A perhaps surprising consequence of lightcone quantization is that we {\it must} introduce a small chiral $\mathbb{Z}_2$-breaking mass $2 i m_\psi \psi \chi$ in order to correctly obtain the spectral functions of the theory.  The reason is that  we integrate out the nondynamical field $\chi$ and it becomes redundant with $\psi$, $\chi \sim \frac{m_\psi}{\partial_-} \psi$.  However, at $m_\psi=0$, $\psi$ and $\chi$ decouple in the free theory, so the $\chi$ field is essentially lost.  This disappearance would seem to conflict with the fact that one can easily think of Feynman diagrams at $m_\psi=0$ where $\chi$ is produced -- for instance, in the fermion loop correction to the $\phi$ mass.  The resolution is that the limits $m_\psi \rightarrow 0$ and $\dmax \rightarrow \infty$ do not commute: as one takes $m_\psi$ smaller, one must take $\dmax$ increasingly large in order for the remaining $\psi$ modes to reproduce the discarded $\chi$.  The role of $\dmax$ in this case is to provide a UV cutoff on $P_+$, through e.g.\ (\ref{eq:freespectrumpsi}), and similar arguments would apply to any other UV regulator in lightcone quantization. 

Perhaps the simplest example where this can be seen is in the two-point function $\< (\psi \chi)(x) (\psi \chi)(y)\>$:
\begin{center}
\begin{equation}
\rho_{\psi \chi}(\mu) = \textrm{Im}\left[ \vcenter{\hbox{\includegraphics[scale=0.3]{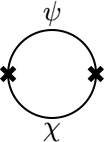}}} \right] \propto \sqrt{1-\frac{4m_\psi^2}{\mu^2}}.
\label{eq:PsiChiLoop}
\end{equation}
\end{center}
The spectral function $\rho_{\psi \chi}$ is nonvanishing even at $m_\chi=0$.  However, the overlap of the operator $(\psi \chi)(x)$ with any two-fermion state $[\psi \psi]_\ell$ is proportional to $m_\psi$, since $\chi$ only creates $\psi$ modes through the equations of motion $\chi \sim \frac{m_\psi}{\partial_-} \psi$.  Naively, there is a contradiction here, because the spectral function at $m_\psi=0$ is a nonvanishing function that is a sum over terms that each individually vanish.  The resolution is that the order of limits $m_\psi \rightarrow 0$ and $\Delta_{\rm max} \rightarrow \infty$ do not commute.  
To understand this discontinuity, consider what happens at $\Delta_{\rm max}=\infty$. In this case, the spectral function is
\be
\rho_{\psi \chi}(\mu) \propto \int_0^1 \frac{dx}{x^2(1-x)^2} | \< \psi \chi | [\psi \psi]_x\>|^2 \delta(\mu^2 - \frac{m_\psi^2}{x(1-x)}).
\ee
Here, $x$ is a momentum fraction of an individual $\psi$ in the two-$\psi$ state, and the factor $x^{-2}(1-x)^{-2}$ in the measure is from the norm of the $\psi \psi$ states. The overlap  of the operator $(\psi \chi)(0)$ with the states $[\psi \psi]_x$ is $\< \psi \chi |  [\psi \psi]_x\> = x^{-1} - (1-x)^{-1}$, and inserting this in the above expression for $\rho_{\psi \chi}$ we obtain the correct answer (\ref{eq:PsiChiLoop}).  However, it is manifest that as $m_\psi$ is taken to be smaller, the contribution to the $\delta$-function comes from smaller $x$, where the energy $m_\psi^2/(x(1-x))$ of the two-particle state is much larger than the mass $m_\psi$.    For finite truncation, then, the problem is clear: if the mass $m_\psi$ is too small, then such states are above the truncation (as for instance one can see from (\ref{eq:freespectrumpsi}).  Consequently, for any finite truncation level, it is necessary to break the chiral $\mathbb{Z}_2$ symmetry by at least a small amount with a fermion mass term.

\subsubsection{Perturbation Theory}
\label{sec:perturbation}

Next we consider how the truncated theory behaves at weak coupling $\bar{g}\ll 1$, where we can use perturbation theory in the coupling $g$.   The UV cutoff is determined by the truncation as described in the previous subsection, and the resulting UV regulator is quite different from more standard regulators. For one, it treats $p_-$ and $p_+$ on different footings.  Moreover, lightcone energy for a massive particle is $\sim m^2/p_-$, which is inversely proportional to $p_-$ and therefore a UV cutoff also acts as an IR cutoff on $p_-$. 

This aspect of the lightcone regulator leads to some perhaps surprising differences in the UV divergences compared to more standard regulators.  Because the theory is super-renormalizable, divergences arise only at low loop order.  Consider the one-loop divergence of the fermion and boson masses:

\begin{center}
\includegraphics[width=0.75\columnwidth]{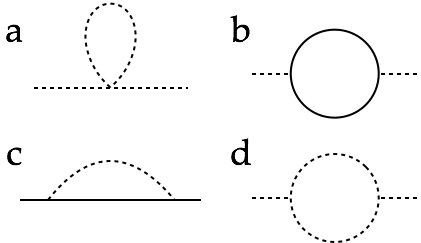} 
\end{center}

 With a standard supersymmetry-preserving regulator, the divergences from diagrams $a$ and $b$  cancel against each other, whereas diagrams $c$ and $d$ are finite.  However, in a Hamiltonian formulation where one computes $P_+$ matrix elements directly (instead of using $Q_+^2$), one usually normal orders the quartic $\phi^4$ interaction, so diagram $a$ is set to zero.  Consequently, the divergence in $b$ does not get canceled and the scalar mass receives a divergent correction.  This correction to $m^2 \phi^2$ breaks the $Q_+$ symmetry, since $Q_+$ relates quadratic, cubic, and quartic terms in the Lagrangian (\ref{eq:PplusLag}) and the cubic and quartic terms do not receive divergent corrections.  In addition, the SUSY enforced relation between the various terms also enforces that the $\mathbb{Z}_2$ is broken spontaneously.  Thus the correction to the scalar mass also breaks $\mathbb{Z}_2$ explicitly.  
Naively, the scalar mass divergence also breaks the $Q_-$ symmetry, which relates the scalar and fermion masses.  But,   a perhaps surprising consequence of the lightcone truncation is that the fermion mass correction from diagram $c$ is also divergent.  Normally, such a divergence is forbidden by chiral symmetry, but note that in the lightcone action (\ref{eq:PplusLag}) with $\chi$ integrated out, the mass term $\sim m^2 \psi \frac{1}{\partial_-} \psi$ is actually {\it invariant} under the $\psi \rightarrow -\psi$ symmetry!  In second-order old-fashioned perturbation theory for the single-fermion energy,  in the continuum ($\Delta_{\rm max}=\infty$) limit, the fermion mass shift is  
\be
\delta m^2 \propto \int_0^1 \frac{dx}{x(1-x)^2}  \frac{|V_{\psi, [\phi \psi]_x}|^2}{m^2 - m^2_{[\phi \psi]_x}}, 
\ee
where $x$ is the momentum fraction of $\phi$ in the $\phi \psi$ intermediate state.
The numerator of this integrand is the matrix element squared $|V_{\psi, [\phi \psi]_x}|^2 = g^2(2-x)^2$ for $\psi \rightarrow \phi \psi$, the denominator is the zero-th order mass-squared difference $\Delta P_+  = P^\psi_+ - P^{\phi \psi}_+ = m^2 - \frac{m^2}{x(1-x)}$, and the factor $x^{-1}(1-x)^{-2}$ is the measure from the norm of the $\phi \psi$ states.  The integral is logarithmically divergent at $x \sim 1$.  The lightcone energy $P_+$ of the two-particle state is $\sim \frac{m^2}{x(1-x)}$, so a UV cutoff $\Lambda_+$ on $P_+$ is also an IR cutoff on small $1-x$, and the log divergence in $x$ becomes $\log \Lambda_+$.  As before,  at finite $\dmax$ the truncation itself sets a UV cutoff $\Lambda_+\sim \dmax m$. 
The upshot, which we have verified numerically, is that both the fermion and scalar mass receive a one-loop correction of the form
\be
\delta m^2 \sim g^2 \log \dmax,
\ee
and moreover the $Q_-$ symmetry enforces that the divergence is the same for both.

Now  let us discuss the status of these divergences when we use $Q_+^2$ to construct the Hamiltonian. As discussed previously, although $Q_+^2=P_+$ at infinite truncation, there is a difference between truncating $Q_+^2$ vs truncating $Q_+$ and then squaring it.  Crucially, in the latter case, diagrams such as $a$ are not discarded by normal-ordering $\phi^4$.  Rather, $\phi^4$ is obtained by ``exchanging a fermion'' between two factors of $\phi^2$ when we compute $Q_+^2$.  Since $\phi^4$ is not normal-ordered in this case, diagram $a$ again produces a divergence that can cancel against the divergence in diagram $b$, and in fact we expect that it must cancel since the construction $P_+\equiv Q_+^2$ manifestly preserves the $Q_+$ symmetry that relates the (finite) cubic and quartic diagrams to the quadratic diagram.  This expectation will be demonstrated in the numerics in later sections through the fact that we see only very weak dependence on $\dmax$ of the mass shift at weak coupling.  

\begin{figure}[h!]
\centering
\includegraphics[width=0.95\columnwidth]{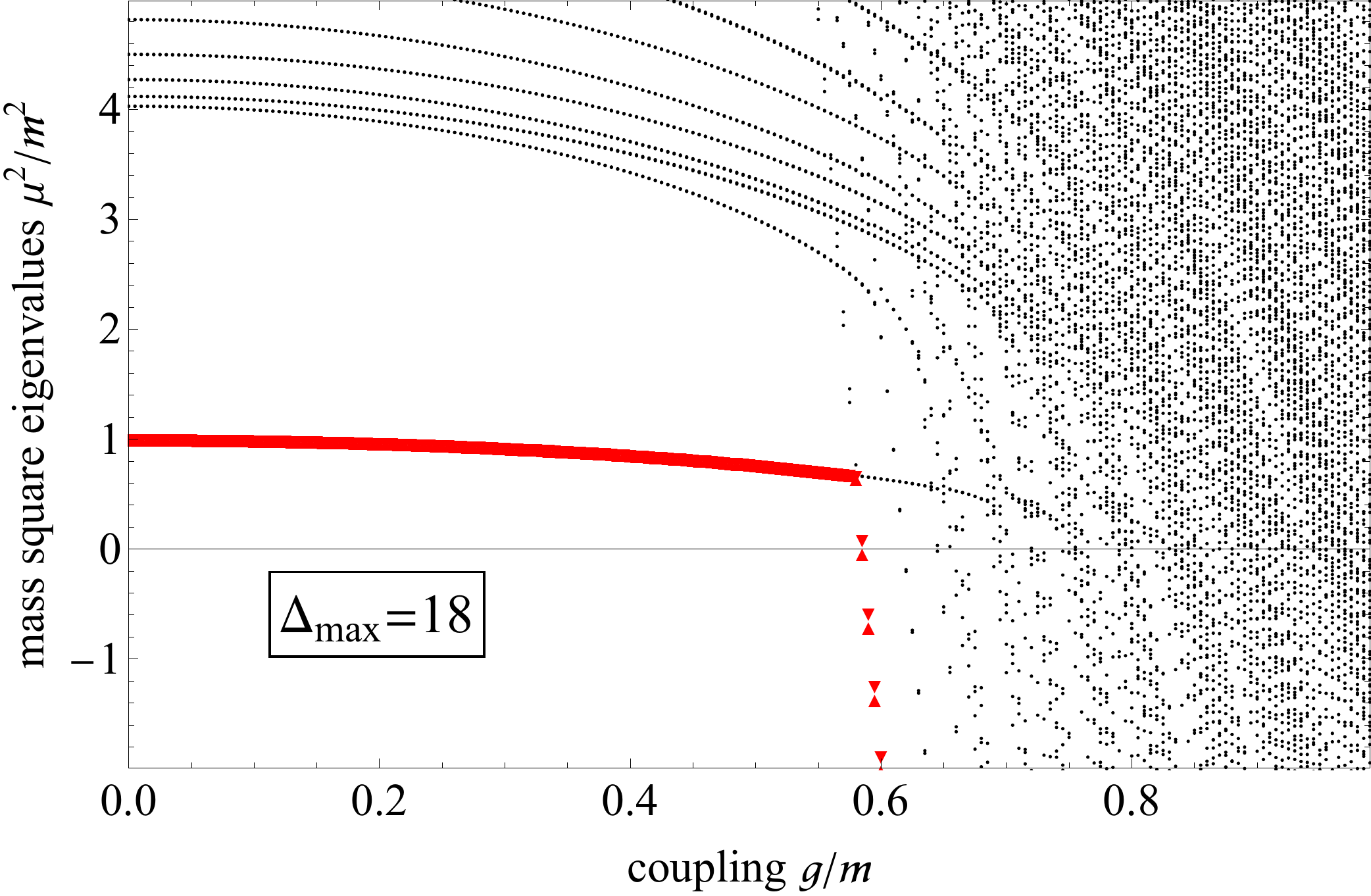}
\caption{\label{fig:lifeIsHardWithPPlus}
Life is hard when using $P_+$ without additional tuning. The eigenvalues of $P_+$ at $\Dmax=18$ are shown above, with the two lowest states highlighted in red. They are (almost) exactly paired by the supercharge $Q_-$. Near $\bar{g} \approx 0.6$, states come from the top of the spectrum and rapidly go to negative mass squared eigenvalues. This sharp drop in the smallest eigenvalue is characteristic of a first order phase transition, which we reach before the critical point.
}
\end{figure}

The fact that the mass does not receive a counterterm in the $P_+ \equiv Q_+^2$ construction is remarkably useful.  For one, it means that physical predictions at different $\dmax$ can be directly compared as a function of coupling $\bar{g}$ without having to compensate for the counterterm.  It also means that the mass is not renormalized and therefore there is a hope of extracting anomalous dimensions from the gap as a function of coupling.  Moreover, we do not have a full understanding of possible nonlocal counterterms that may be induced in the subleading finite part of the divergent diagrams. Finally, if we were to construct matrix elements for $P_+$ directly, then as discussed above, at large $\dmax$ and large coupling we would have to include a counterterm and fine tune it to restore both SUSY and the $\mathbb{Z}_2$ symmetry, which is cumbersome.  Without the counterterm, there would be no guarantee that we would reach the critical point simply by scanning over $\bar{g}$.  Indeed, as we show in Fig.\ \ref{fig:lifeIsHardWithPPlus}, in the construction based on computing $P_+$ directly without any additional tuning, we appear to hit a first-order phase transition before reaching the critical point. For these reasons, our numeric analysis at strong coupling will use the $Q_+^2$ construction unless stated otherwise.

\subsection{\texorpdfstring{$H_{\rm eff}$}{Heff}}

One final important feature of perturbation theory is that it provides a regime where we can explicitly compute the effect of zero modes discarded by lightcone quantization.  One can think of lightcone quantization as `integrating out' $p_-=0$ zero modes, potentially leaving behind additional terms in a new effective lightcone Hamiltonian $H_{\rm eff}$. To all orders in perturbation theory, 
a scalar with a $\lambda \phi^4$ interaction generates a shift in the mass proportional to $\lambda \< \phi^2\>$ \cite{Burkardt,BurkardtEtAl,Heff}, and \cite{Borel} argued that the nonperturbative shift in the mass could be obtained from the perturbative shift by comparing the Borel-resummation of the perturbation series for the mass in equal-time and lightcone.  The basic point for $\phi^4$ theory is that lightcone quantization does not just discard the `normal-ordering' piece (i.e.\ diagram $a$ above) of the mass shift from $\phi^4$, it also discards all diagrams where additional interactions are added on the loop in diagram $a$.  Perturbatively, the full set of such diagrams is proportional to $\< \phi^2\>$.  However, in the supersymmetric theory,
 in the SUSY-preserving phase $\bar{g}< \bar{g}_*$, 
there are no corrections to $\< \phi^2\>$ as a simple consequence of SUSY preservation, since from (\ref{eq:FirstSPotl}) we have $\<W'(\phi)\> = h+ \frac{g}{2} \<\phi^2\> = 0$~\cite{Burkardt:1998rk}.\footnote{Note that this argument fundamentally relies on a perturbative expansion around $g=0$, and so cannot be trusted at $\bar{g}> \bar{g}_*$ on the other side the phase transition.}  This fact will be important when we extract the critical exponent $\nu$ from the mass gap as a function of $\bar{g}$, because a nonvanishing $H_{\rm eff}$ could spoil the relation (\ref{eq:criticalExponent}) (and in fact does spoil it for nonsupersymmetric $\phi^4$ theory \cite{Borel,Phi4Paper}).

\section{\texorpdfstring{$\mathbb{Z}_2$}{Z2}-breaking Phase and TIM}
\label{sec:massive}

Next we move on from the perturbative regime to study the SGNY theory numerically at strong coupling in our truncation framework. 
For each value of the dimensionless coupling $\bar g\equiv g/m$, we take our Hamiltonian (\ref{eq:Hamiltonian})  truncated at $\dmax$ 
and numerically find its eigenvalues and eigenvectors. 
The mass spectrum of the SGNY theory as a function of $\bar g$ are just these eigenvalues, and spectral functions can be computed from the eigenvectors using (\ref{eq:GenSpecFunc}).

Because the vacuum energy is automatically set to zero in LCT, the lowest mass eigenvalue of our Hamiltonian is the mass gap of the theory. At weak coupling, the gap is approximately just the bare mass, and decreases with increasing coupling $g$ until it  closes at the critical coupling $\bar g_*$, where the IR of the theory is the TIM critical point. Slightly away from the critical point in either direction,  the IR theory is  TIM with a relevant SUSY-preserving deformation $\epsilon'$:
\be
	\CL_{\rm UV} 
	\Rightarrow \CL_{\rm TIM} - \lambda \epsilon^\prime \, .
\ee
where the arrow represents the RG flow from the UV to the IR. In this section, we will focus on the range $\bar{g}< \bar{g}_*$, where the theory breaks the $\mathbb{Z}_2$ symmetry spontaneously and by dimensional analysis $\lambda$ is proportional to $m_{\rm gap}^{4/5}$.  We will discuss the range $\bar{g} > \bar{g}_*$, where the theory breaks SUSY spontaneously and the gap remains zero, in  Section \ref{sec:susy-breaking-phase}.

\subsection{Mass Spectrum of Interacting Theory}
\label{sec:mass-spectrum-interacting}

\begin{figure}[ht!]
\centering
\includegraphics[width=0.95\columnwidth]{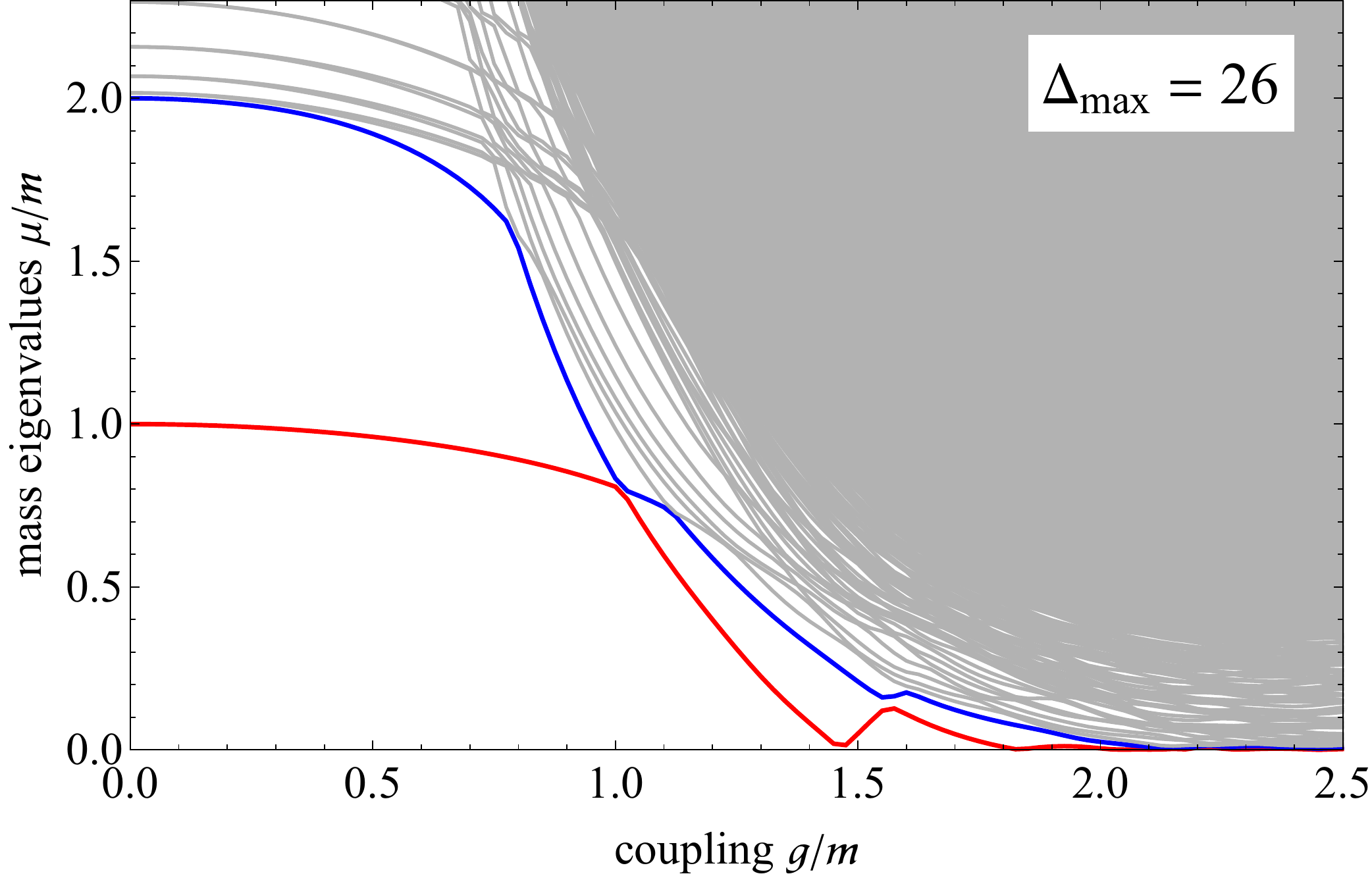}
\caption{\label{fig:spectrum}
	Mass eigenvalues of the truncated Hamiltonian $Q_+^2$ as a function of the coupling $\bar{g}$. We truncate the basis at $\dmax = 26$, which includes 40824 states. Each mass eigenvalue is exactly 2-fold degenerate due to supersymmetry. We highlight the lowest two mass eigenvalues at each $g$ with red and blue solid curves. The rest of the eigenvalues are gray solid curves. The higher states become very dense and their curves fill the upper right region.
}
\end{figure}

We begin with the simplest observable, the mass spectrum as a function of the coupling, shown in Fig.\ \ref{fig:spectrum}. All states come in exact pairs due to the $Q_+$ symmetry. The lowest eigenstate is the mass gap (shown in red).
At zero coupling, the gap is just the mass term $m$, and it decreases as the coupling $\bar g$ gets stronger.  For weak coupling,  the next state (blue) is the threshold of two-particle states, above which we see a near-continuum of two-particle states that is discretized due to the truncation.  These states come in approximately degenerate sets of 4 (pairs of pairs), due to the approximate $Q_-$ symmetry.

Near $\bar g \approx 1$ the gap turns down more rapidly as the one-particle state and two-particle continuum eigenvalues collide.\footnote{  This feature is somewhat surprising. 
A possible explanation is that at weak coupling, the lowest states are particles, whereas near the TIM critical point, the lowest states are massive ``kinks'' which do not form bound states. In the intermediate regime, there must be a transition between bounded particle states to a continuum of unbounded kinks.  It is possible that the bound states remain stable until they cross over the kink states.
}  For now, note that as we continue increasing the coupling eventually the gap closes at $\bar g_* \approx 1.5$.  Near $\bar g_*$ we have a prediction, (\ref{eq:criticalExponent}), that the mass gap closes as the critical exponent $\nu$ of TIM. We will study this prediction numerically in Section \ref{sec:criticalExponent}. Note that although we expect the spectrum to be continuous at the critical point, in our figure the first and second eigenvalues do not reach zero at the same coupling.  This is due to truncation effects, and we expect that as the results converge to their infinite $\dmax$ limit, the higher eigenvalues will close at the same coupling as the lowest one. 

Finally, for $\bar g > \bar g_*$ the gap fluctuates near zero, until $\bar g \approx 2$ where the spacing become invisible.  This result is in agreement with the expectation of a massless Goldstino when SUSY is spontaneously broken.

\begin{figure}[ht!]
\centering
\includegraphics[width=0.95\columnwidth]{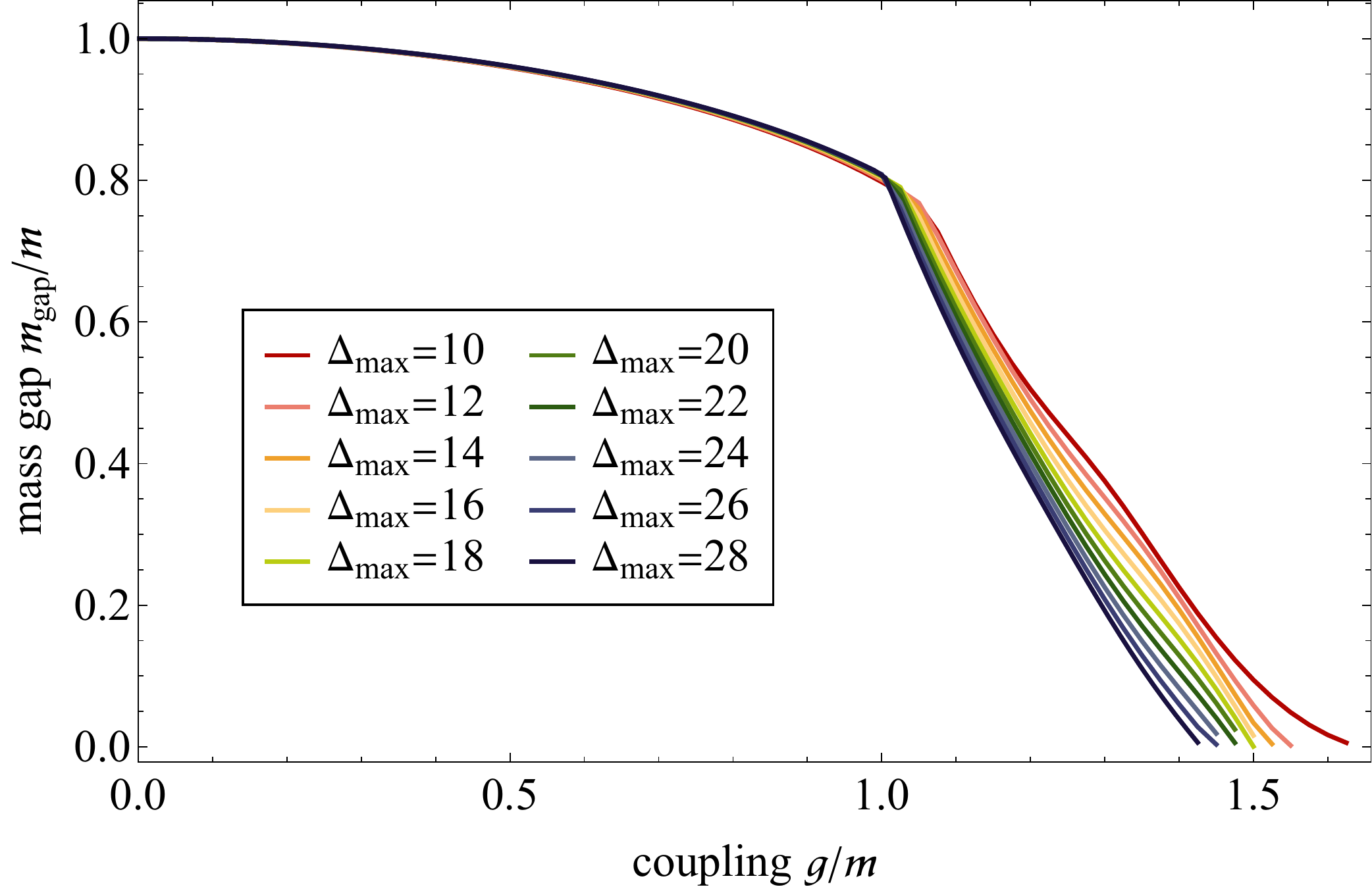}
\caption{\label{fig:gap}
	The mass gap $\mgap$ as a function of the coupling $\bar{g}$, at different $\dmax$. 
	At each $\dmax$ the mass gap is a smooth function of $\bar g$, with one kink at $\bar g \approx 1$ where the continuum collides with the single particle state. The mass gap before the collision is almost the same for all $\dmax$. As $\dmax$ increases (from red to blue), the coupling $\bar g_0$ at the collision moves to the left. The critical coupling $\bar g_*$ where the gap closes also shifts to the left. The shape of the curve between $\bar g_0$ and $\bar g_*$ deforms for small $\dmax$ and stabilizes for large $\dmax$.
}
\end{figure}

Fig.\ \ref{fig:gap} shows how the mass gap converges as we increase the truncation level $\dmax$. The gap has converged well in the small coupling regime, where the gap is set by the single particle state which is well-separated from the more energetic continuum. By contrast, for larger couplings beyond where  the continuum states cross the one-particle state, the gap is still visibly changing as we increase $\dmax$ even at the highest truncation $\dmax=28$. This means it may be tricky to extract physical data from an individual gap value at fixed $\bar g$.
However,  the general shape of the function seems to have stabilized and behave much better than individual mass eigenvalues. 

\subsection{Critical Exponent}\label{sec:criticalExponent}

Next, we zoom in to the vicinity of the critical point, on the gapped side. In the IR, the theory flows to TIM with a relevant deformation $\epsilon^\prime$. Dimensional analysis demands that the gap should vanish as a power law of the small parameter $(g_* - g)$
\begin{align}\tag{\ref{eq:criticalExponent}}
	\mgap \propto (g_*-g)^{\nu}, ~~~~~\nu = \frac{5}{4} \, ,
\end{align}
where $\nu$ is the critical exponent. In this section we study the critical exponent quantitatively from our truncation data.
Recall in Fig.\ \ref{fig:spectrum} the mass gap turns down sharply after the continuum collides with the single particle state. We will choose to fit to this region. For each $\dmax$, we take the gap function between the collision $\bar g_0$ and $\bar g_*$, and fit this function to a power law.

\begin{figure}[h!]
	\centering 
	\includegraphics[width=0.85\columnwidth]{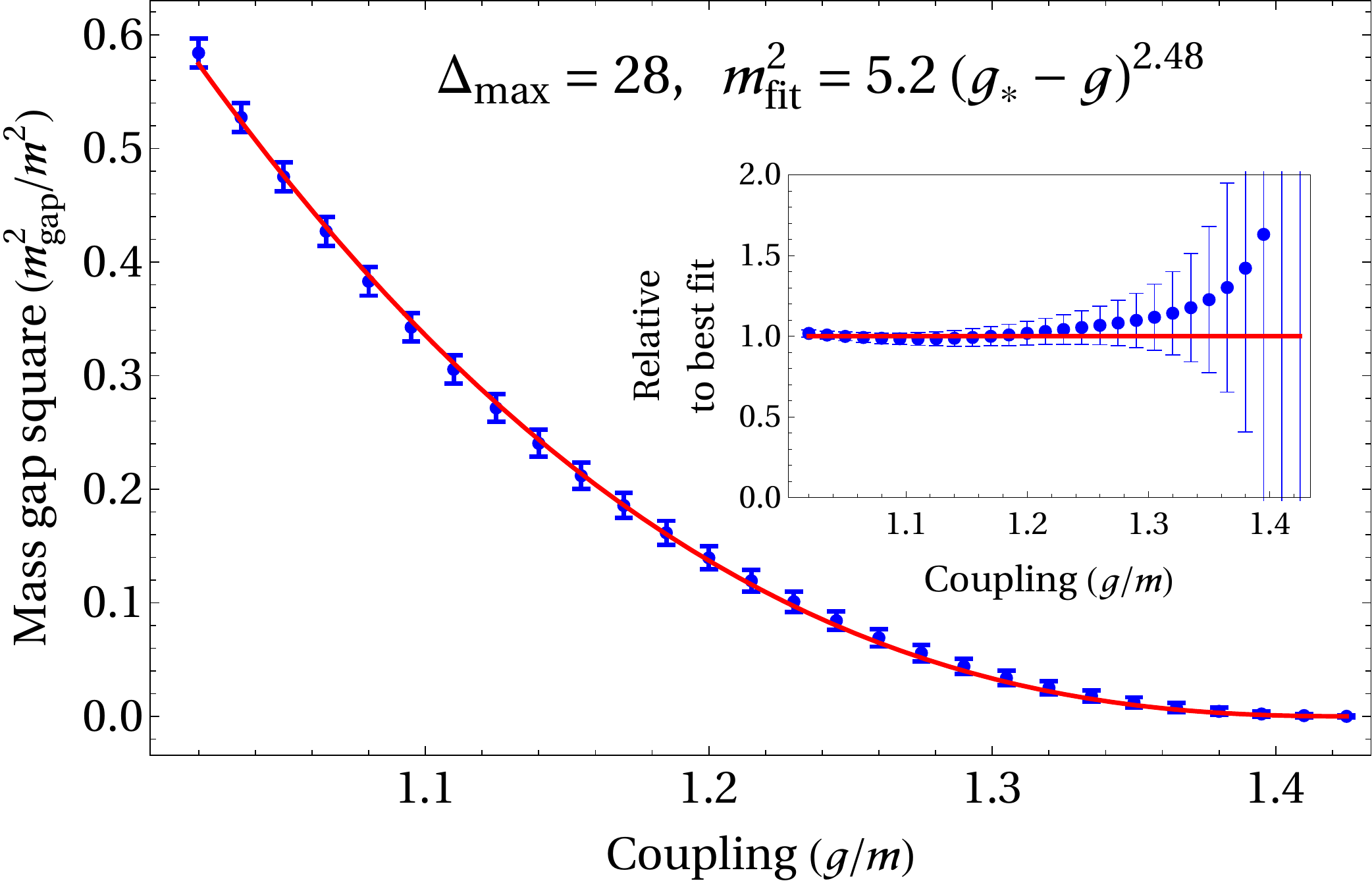} 
	\caption{\label{fig:powerLawFit} 
	The power law fit to $\mgap^2$ as a function of $\bar g$ near the critical coupling $\bar g_*$ on the massive side. The blue dots are the $(\bar g_i, m_{{\rm gap},i}^2)$ data points obtained from diagonalizing the Hamiltonian $Q_+^2$ at coupling $\bar g_i$. The magnitude of the error bar on the points are defined as the change of $\mgap$ magnitude between $\dmax = 26$ and $\dmax = 28$ fixing $\bar g_i$.  
	The red curve is the best fit. 
	The inset plot shows the same data and fit, normalizing the best fit $m_{\rm fit}^2$ to 1. 
	}
\end{figure} 

\begin{figure}[h!]
\centering
\includegraphics[width=0.95\columnwidth]{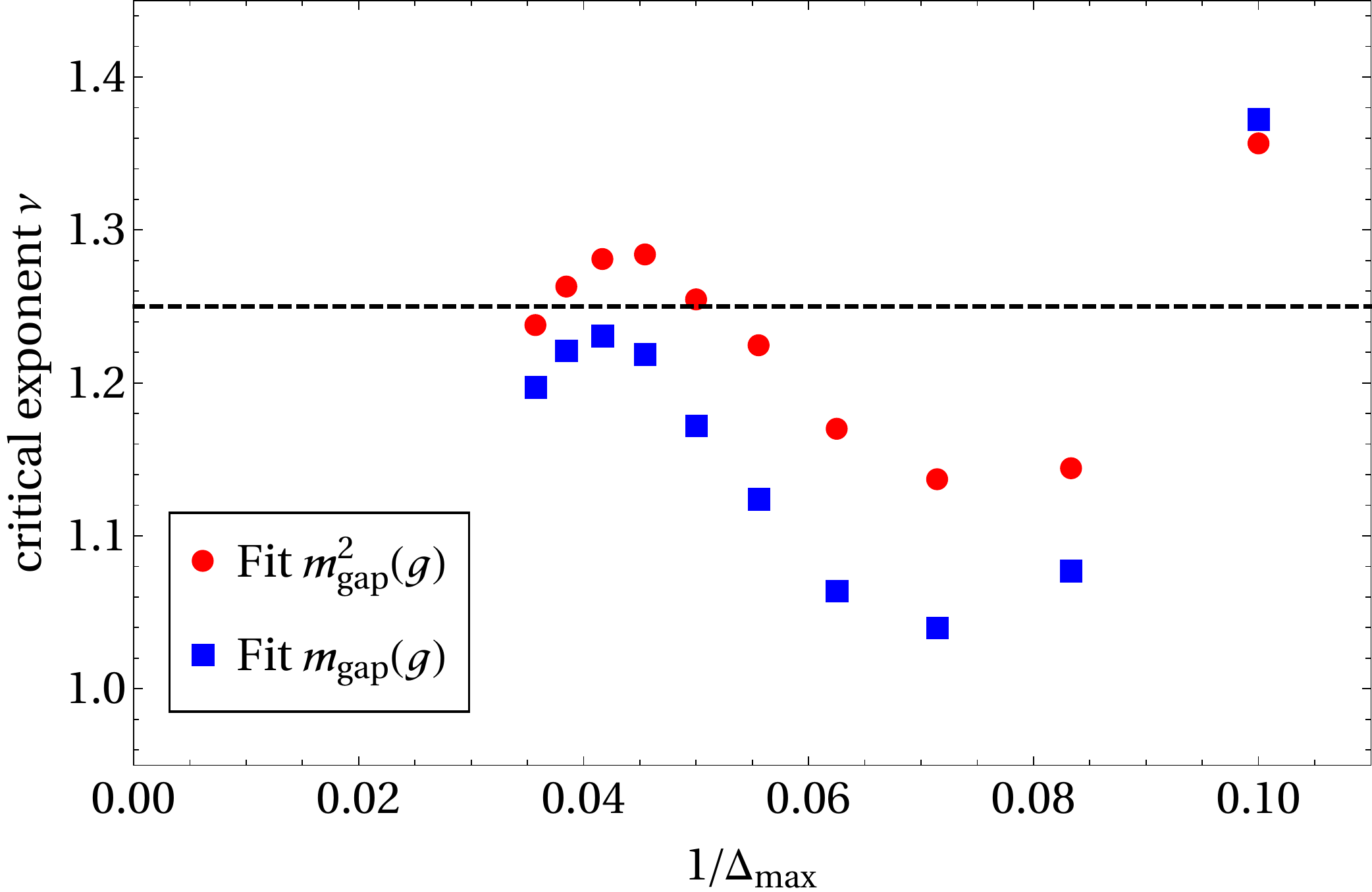}
\caption{\label{fig:criticalExponent} 
The numerical fit to the critical exponent $\nu$ of the SUSY Yukawa theory near the critical coupling. The theoretical expectation (black dashed line) from TIM is $\nu = \frac{5}{4}$. The best fit values (red dots) oscillate around the theoretical prediction and approch it as $\dmax$ increases. The blue squares are the critical exponents extracted from the same data using the same procedure but fitting to $\mgap$ instead of $\mgap^2$.
}
\end{figure} 

Fig.\ \ref{fig:powerLawFit} shows an example fit to the mass gap data.
We do the fit at each $\dmax$ and extract the critical exponent.
The results are summarized in Fig.\ \ref{fig:criticalExponent}.
We sketch the procedure as follows. 
First, for each $\dmax$ we scan the coupling $\bar g$ with a small step size and diagonalize the Hamiltonian $Q_+^2$ at each $\bar g_i$, obtaining the mass gap $ m_{{\rm gap},i}$. From the data $\{(\bar g_i, m_{{\rm gap},i})\}$ we locate the beginning $\bar g_0$ and the end $\bar g_*$ of the fit range, and restrict the data to $\bar g_0 \leq \bar g_i \leq \bar g_*$. We then fit the data to 
\begin{align}
	m_{\rm fit}(\bar g_i)^2 = a (\bar g_* - \bar g)^{2\nu}
\end{align}
with $\bar g_*$ fixed and varying $a$ and $\nu$. We take the $\nu$ from the best fit that minimizes the total least-square-error 
\begin{align}
	\sum_i \pr{m_{\rm fit}(\bar g_i)^2 - m_{{\rm gap},i}^2}^2 \, .
\end{align}
Specifically, 
we choose to fit $\mgap^2$ as a function of $\bar g$. However, there is some ambiguity in this choice, and one could instead choose to fit, say, $\mgap$ as a function of $\bar{g}$ in order to extract $\nu$.  As a measure of the uncertainty arising from this ambiguity, in Fig.\ \ref{fig:criticalExponent} we show the critical exponent extracted from both fitting $\mgap^2$ and $\mgap$. Note that as $\dmax$ increases, the extracted value of $\nu$ from these two fits becomes closer, indicating that the uncertainty of the fit is also shrinking as $\dmax$ increases.

Our best numerical result, at $\dmax=28$, is 
\begin{align}
	\nu = 1.24 \pm 0.05 \, .
\end{align}
The central value is obtained from the best fit of $\mgap^2(g)$ at $\dmax=28$. The uncertainty is estimated as the difference between the best fit value at $\dmax=28$ and $\dmax=22$. We use this difference as our uncertainty because Fig.\ \ref{fig:criticalExponent} shows that the measured $\nu$ oscillates with a shrinking amplitude and $\dmax=22$ is the nearest peak. The difference between different ways of fitting the mass gap is also of the same magnitude. The result is clearly consistent with the TIM theoretical expectation $\nu=1.25$.

It is worth mentioning that  $\nu=1$ is ruled out by our error estimates. 
The reason this is interesting is that one could easily imagine obtaining $\nu=1$ for completely unphysical reasons, due to the fact that we are studying a truncated system.  That is, since $\mgap$ is an eigenvalue of the finite dimensional matrix $Q_+$, in general it should be an analytic function in the parameter $g$. In fact, if we get too close to the critical point, the gap as a function of coupling can always be series expanded around $g_*$ where the leading power of $g_*-g$ must be an integer at finite truncation. 
As $\dmax$ increases, however, this linear region shrinks and  we can more accurately obtain the critical exponent.

\subsection{Zamolodchikov \texorpdfstring{$C$}{C}-Function}
\label{sec:CFunc}

\begin{figure*}[t!]
\centering
\includegraphics[width=0.95\textwidth]{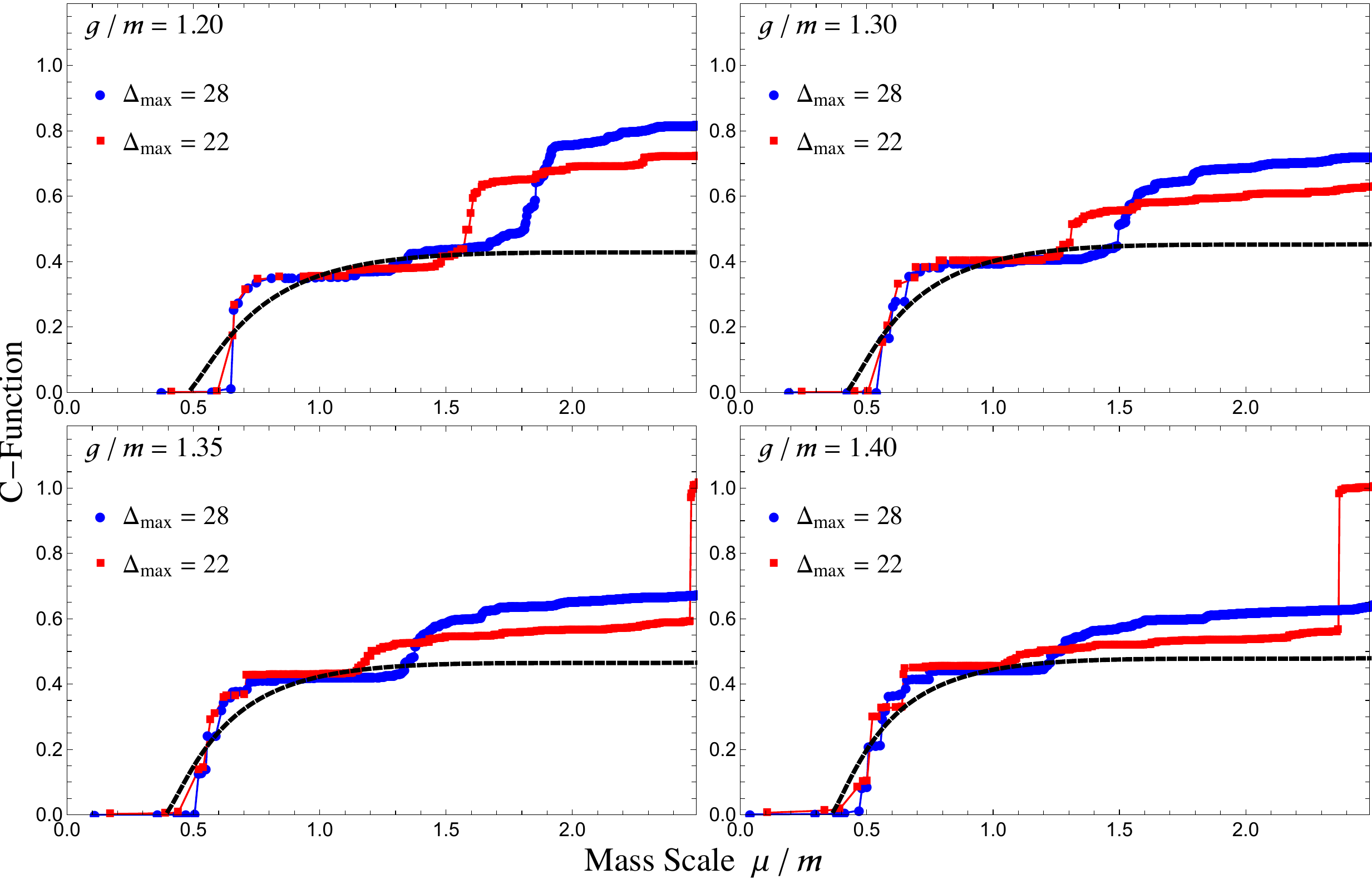}
\caption{\label{fig:CFuncGrid} 
Zamolodchikov $C$-function at different values of $\bar{g}$ near the critical point for $\Dmax = 28$ (blue dots) and $\Dmax=22$ (red squares). The truncation data for SGNY is compared to the IR theoretical prediction for TIM from eq.~\eqref{eq:CFuncTIM} (black dashed line), which includes the effects of the relevant deformation $\epsilon'$ as well as the leading higher-dimensional operator $\p^2\epsilon'$. The value for $\mgap$ for the TIM prediction was obtained by fitting to the $\Dmax=28$ data for the integrated $C$-function (see appendix~\ref{app:doubleInts}) at each value of $\bar{g}$, while the value $\Lambda \approx 10m$ was obtained by specifically fitting to the data at $\bar{g}=1.35$.
}
\end{figure*} 

As mentioned in section~\ref{sec:setup}, we can use the eigenstates $|\mu_i\>$ of the truncated Hamiltonian to compute dynamical observables, such as the spectral functions of local operators $\rho_\Ocal(\mu)$. However, because spectral functions formally correspond to a sum of delta functions, in practice it is simpler to study integrated spectral function, which correspond to the cumulative overlaps of the eigenstates $|\mu_i\>$ with individual local operators,
\be
I_\Ocal(\mu) \equiv \int_0^{\mu^2} d\mu'^2 \rho_\Ocal(\mu') = \sum_{\mu_i \leq \mu} \big|\<\Ocal(0)|\mu_i\>\big|^2.
\ee
One particularly useful operator to study in 2D is the stress-energy tensor component $T_{--}$, whose integrated spectral function corresponds to the Zamolodchikov $C$-function~\cite{Zamolodchikov:1986gt,Cappelli:1990yc},
\be
C(\mu) \equiv \fr{12\pi}{P_-^4} I_{T_{--}}(\mu).
\label{eq:CfuncDef}
\ee
Famously, this monotonically increasing function of $\mu$ interpolates between the central charges of the IR and UV fixed points.

For the SGNY theory, we can construct the $C$-function by computing the overlaps of the mass eigenstates with the UV operator
\be
T_{--} = (\p_-\phi)^2 + i\psi\p_-\psi.
\ee
In the SUSY-preserving phase, we generically expect $C(\mu)$ to start at $c_\IR=0$ (since the theory has a mass gap), then increase to eventually reach the UV value $c_\UV = \fr{3}{2}$ as $\mu \ra \infty$. Near the critical point, though, we expect the IR behavior of $C(\mu)$ to match that of the $\epsilon'$-deformed TIM. This provides us with a concrete prediction for the low-energy behavior of the $C$-function as $\bar{g} \ra \bar{g}_*$,
\be
C(\mu) \propto \fr{\mgap^{\frac{8}{5}}}{\mu^4} \rho_{\epsilon'}(\mu) \qquad (\bar{g} \ra \bar{g}_*, \, \mu \ll g).
\ee
Fortunately, the deformation of TIM by $\epsilon'$ is integrable, so the theoretical prediction for $C(\mu)$ can be computed analytically (see appendix~\ref{app:TIMFormFactors} for more details).

However, if $\mgap$ is not sufficiently separated from the UV scale $g$, there are additional corrections due to higher-dimensional TIM operators, specifically those which preserve SUSY. The leading correction comes from the descendant $\p^2 \epsilon'$, with the contribution suppressed by some scale $\Lambda$,
\be
S_\IR \approx S_\TIM + \int dx \left( \mgap^{\fr{4}{5}} \epsilon' - \fr{1}{\Lambda^{\fr{6}{5}}} \p^2 \epsilon' + \cdots \right).
\ee
Including this leading correction, we therefore obtain the full TIM prediction
\be
C_\IR(\mu) \propto \fr{\mgap^{\fr{8}{5}}}{\mu^4} \left( 1 - \fr{2\mu^2}{\mgap^{\fr{4}{5}} \Lambda^{\fr{6}{5}}} + \fr{\mu^4}{\mgap^{\fr{8}{5}} \Lambda^{\fr{12}{5}}} \right) \rho_{\epsilon'}(\mu).
\label{eq:CFuncTIM}
\ee

Figure~\ref{fig:CFuncGrid} shows the truncation results for $C(\mu)$ at $\Dmax=28$ (blue dots) at four different values of $\bar{g}$ near $\bar{g}_*$, compared with the TIM prediction from eq.~\eqref{eq:CFuncTIM} (dashed black line). The TIM prediction has two free parameters: $\mgap$ and $\Lambda$. For each plot, the value of $\mgap$ was obtained by fitting to the data.\footnote{Specifically, the value was extracted by fitting to the integral of the $C$-function, which in practice is a much smoother function (see appendix~\ref{app:doubleInts}).} However, because the scale $\Lambda$ should be proportional to the coupling $g$, its value should not change by much in these four plots. As a simple sanity check, we therefore have fit the value of $\Lambda$ using only the data at $\bar{g}=1.35$, obtaining $\fr{\Lambda}{m} \approx 10$. As we can see, this value of $\Lambda$ still provides excellent agreement with the truncation data in the remaining three plots.

For reference, we have also provided results at $\Dmax=22$ (red squares), in order to convey the level of convergence of the truncation data. At low values of $\mu$, the data largely agrees with the TIM prediction, then begins to deviate as we proceed to higher $\mu$. This deviation towards the UV is physical, and arises because the SGNY model is not identical to the TIM, and only flows to it in the IR. Additionally, we see that the correction due to $\p^2\epsilon'$ is significant, lowering the IR plateau away from the naive value of $c_\TIM = \fr{7}{10}$. Again, this correction is physical, and its size is set by the ratio of $\mgap$ with respect to $g$. If we had infinite IR resolution, such that we could accurately study the theory at energies orders of magnitude below $g$, then this correction would diminish, raising the plateau to the naive TIM value.

At finite truncation, we also see that the $C$-function has a step-like structure, where particular mass eigenstates provide the main contributions to $C(\mu)$, leading to discrete jumps in the function. We can smooth out these steps by integrating a second time, which we show in appendix \ref{app:doubleInts}. Doing such additional integrations not only smooths out the spectral functions, but moreover it decreases the relative error of the truncation result compared to the analytic result.

\subsection{Trace of Stress Tensor}
\label{sec:Trace}

\begin{figure*}[t!]
\centering
\includegraphics[width=0.95\textwidth]{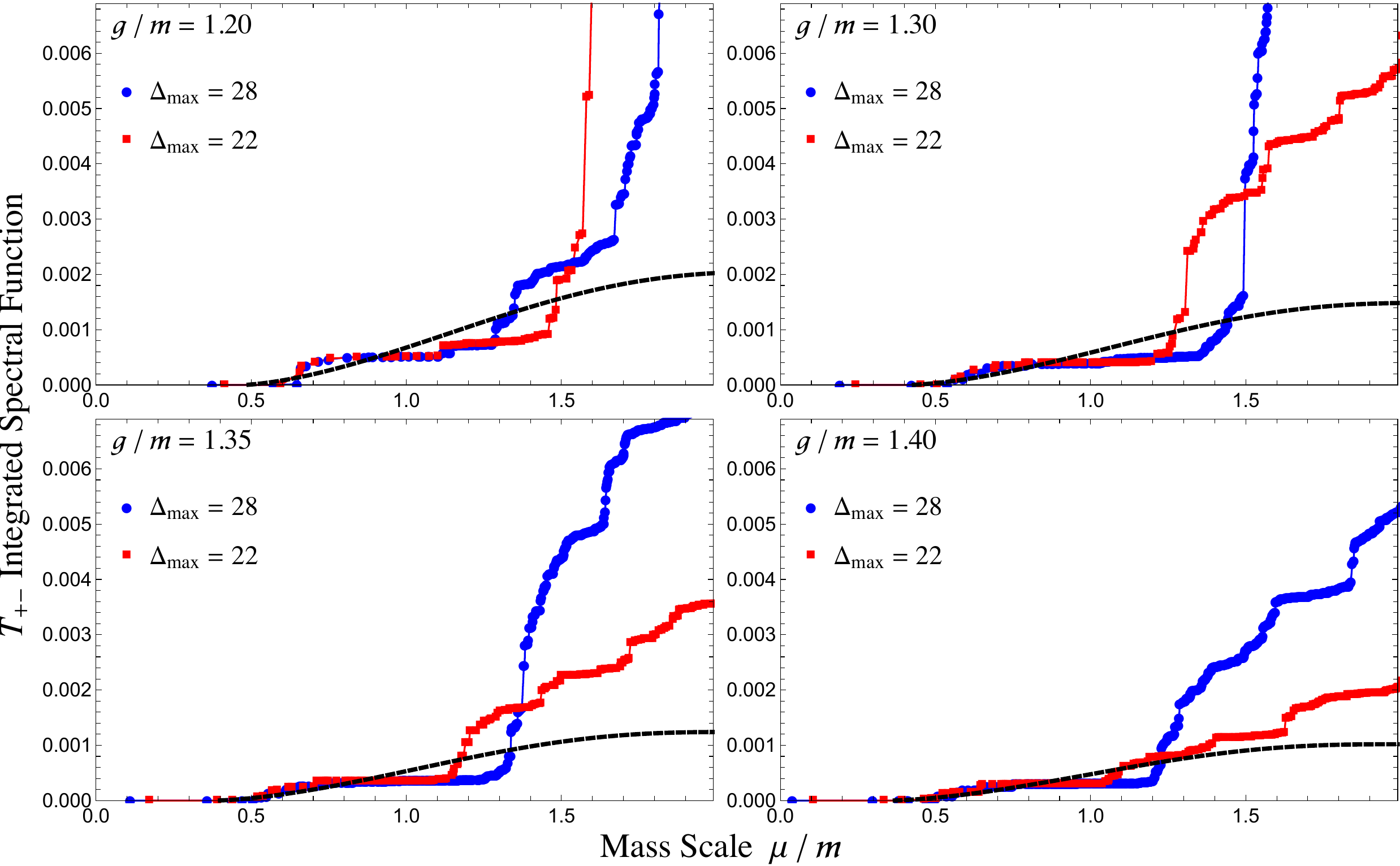}
\caption{\label{fig:TraceGrid} 
Integrated spectral function for the trace of the stress tensor at different values of $\bar{g}$ near the critical point for $\Dmax = 28$ (blue dots) and $\Dmax=22$ (red squares). The truncation data was obtained from the spectral function for $T_{--}$ by using the Ward identity~\eqref{eq:WardId}, and is compared to the IR prediction for TIM (black dashed line). The values for $\mgap$ and $\Lambda$ in the TIM prediction are the same as those in figure~\ref{fig:CFuncGrid}.
}
\end{figure*}

Another operator we can consider is the trace of the stress-energy tensor
\be
T^\mu_{\phantom{\mu}\mu} = 2T_{+-}.
\ee
Technically, the spectral function for $T_{+-}$ is not an independent observable, since it is related to the spectral function of $T_{--}$ via the Ward identity
\be
P_+ T_{--} + P_- T_{+-} = 0.
\label{eq:WardId}
\ee
However, the trace of the stress tensor is still useful, as it must vanish if the theory flows to a CFT in the IR. For the case of the SGNY model, we know that the phase transition is described by TIM, so we therefore expect the IR behavior
\be
T_{+-} \ra 0 \, \, \, \textrm{as} \, \, \, \bar{g} \ra \bar{g}_*.
\ee
More concretely, near the critical point we expect that $T_{+-}$ should match the spectral function of the TIM operator $\epsilon'$, with the leading correction coming from the descendant $\p^2\epsilon'$ (just like in the previous section),
\be\label{eq:TraceTIM}
T_{+-} \propto \mgap^{\fr{4}{5}} \epsilon' - \fr{1}{\Lambda^{\fr{6}{5}}} \p^2 \epsilon' + \cdots
\ee

Figure~\ref{fig:TraceGrid} shows the integrated spectral function for $T_{+-}$ at $\Dmax=28$ (blue dots), compared with the theoretical prediction from TIM (black dashed line). The values for $\mgap$ and $\Lambda$ in the TIM prediction are the same as those used in figure~\ref{fig:CFuncGrid}. As we can see, the spectral function clearly vanishes in the IR as $\mgap \ra 0$, confirming that the critical point is described by a CFT. The deviation from zero also matches the TIM prediction at low energies, eventually deviating as we go to the UV.  As with the $C$-function, we show in appendix \ref{app:doubleInts} that the relative error compared to the continuum prediction can be reduced by integrating the spectral function once more.

\subsection{Universal IR Scale Due to Truncation}
\label{sec:universal-IR}

We have seen that the critical exponent prediction fits well to a range of mass gap values computed numerically from LCT, and that the fit breaks down when we get too close to the critical point. In other words, there is an IR scale beyond which the Hamiltonian truncated at finite $\dmax$ does not have enough resolution.
We expect such a scale since in Hamiltonian truncation we are trying to approximate a Hamiltonian eigenstate in the IR using a finite basis. The convergence usually becomes worse when approaching a fixed point, where the deep IR is very far from the UV.
 In this subsection we would like to probe the IR scale of LCT using a simple scaling ansatz.

\newcommand{\tmgap}{\tilde m_{\rm gap}}
We consider the ratio of the observed mass gap $\tmgap(\dmax)$ and the exact mass gap $\mgap \propto (g_*-g)^{\nu}$. 
At finite truncation, we would like to propose that there is an IR scale $\Lambda_{\rm IR} \sim \dmax^{-\alpha}$, such that below the scale $\mgap \ll \Lambda_{\rm IR}$ the observed mass gap, $\tmgap$, is dominated by a universal function of the dimensionless quantity $\mgap/\Lambda_{\rm IR}$.  Using our assumption for $\Lambda_{\rm IR}$ in terms of $\dmax$, and the behavior of the gap $\mgap$ in terms of the coupling near the critical point, we can write this dimensionless quantity in terms of $\dmax$ and $g$ as
\begin{align}\label{eq:IR-scale}
	\frac{\mgap}{\Lambda_{\rm IR}} \propto \dmax^\alpha (\bar g_*- \bar g)^{\nu} \, .
\end{align}
Above the IR scale, the observed mass gap $\tmgap$ should approach the exact mass gap, so the ratio is constant
\begin{align}\label{eq:universal-above-IR}
	\frac{\tmgap}{(\bar g_* - \bar g)^{\nu}} = {\rm const} \, , ~~~~ \mgap \gg \Lambda_{\rm IR} \, .
\end{align}
However, below the IR scale, the observed gap $\tmgap$ will be increasingly sensitive to IR effects.  If these corrections depend on $\dmax$ only through a universal function of $\mgap/\Lambda_{\rm IR}$, then we can generalize (\ref{eq:universal-above-IR}) to
\begin{align}\label{eq:universal-below-IR}
	\frac{\tmgap}{(\bar g_* - \bar g)^{\nu}} = f\pr{ \dmax^\alpha (\bar g_*- \bar g)^{\nu} }. 
\end{align}

\begin{figure}[h!]
\centering
\includegraphics[width=0.95\columnwidth]{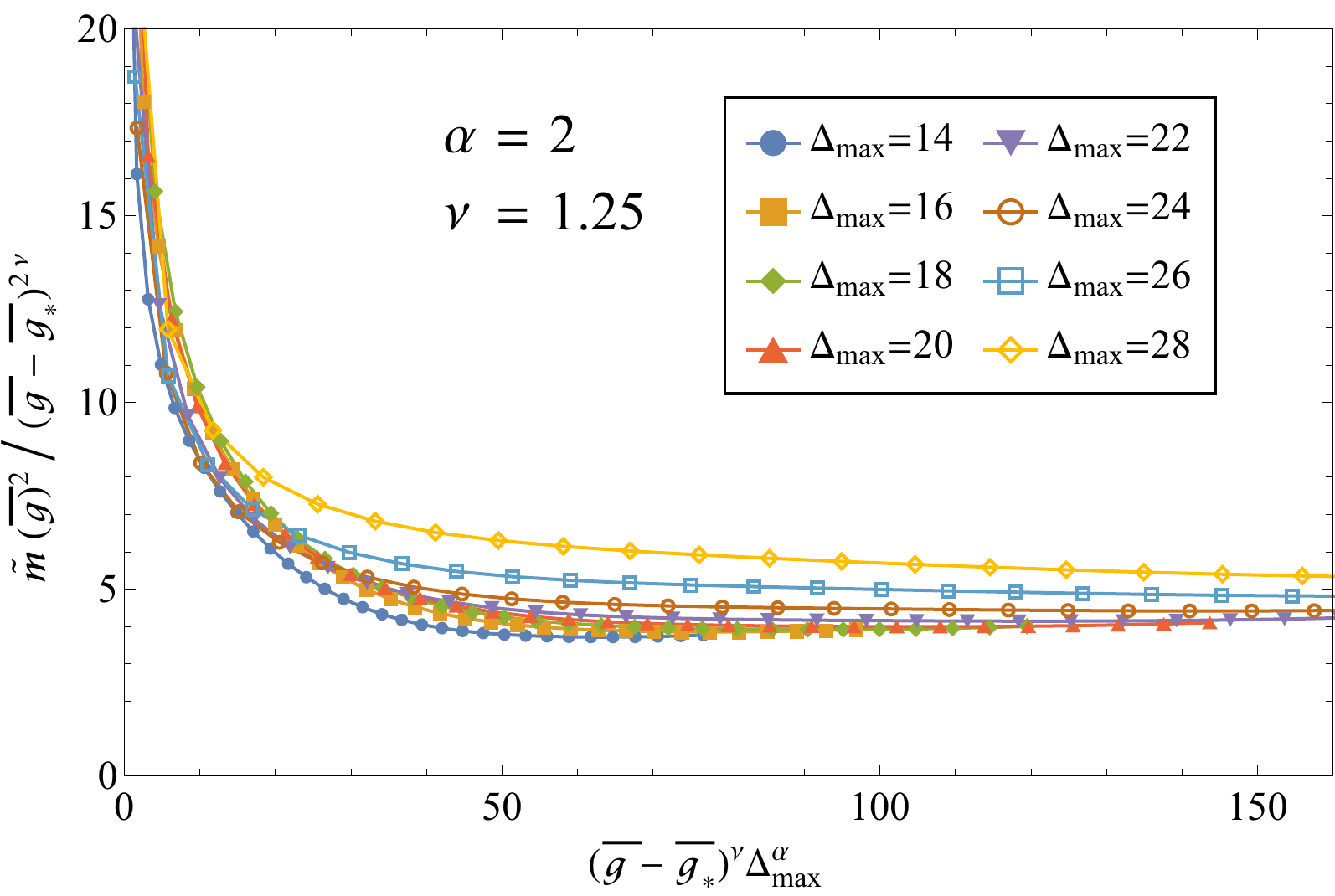}
\caption{\label{fig:universalIR} 
Testing the scaling collapse ansatz of the IR truncation effects. The hypothesis is that the mass gap data at different truncation level $\dmax$ are related by (\ref{eq:universal-above-IR}) and (\ref{eq:universal-below-IR}) above and below the proposed IR scale (\ref{eq:IR-scale}), respectively.
Taking the theoretical value of the critical exponent $\nu = 1.25$ we find from observation that the best parameter that relates different $\dmax$ is near $\alpha\approx2$. 
First, above the IR scale (large $x$-axis), the ratio ($y$-axis) between the mass gap and the power law becomes constant (\ref{eq:universal-above-IR}) for all $\dmax$, indicating that the mass gap is well-fitted by a power law with the correct critical exponent. 
Second, below the IR scale (small $x$-axis) the curves for all $\dmax$ merge to the same behavior, indicating the truncation effect attracts the mass gap to a universal behavior modeled by (\ref{eq:universal-below-IR}) with the parameter $\alpha\approx2$. 
}
\end{figure}  

We check the agreement between the above two equations and our data in Fig.\ \ref{fig:universalIR}. The plot shows the key features that we propose. There is a clear, uniform IR scale on the $x$-axis. To the right, the mass gap is above the IR scale, and the ratio is constant. To the left, the mass gap is below the IR scale, the mass gap deviates from the critical exponent and behaves uniformly across different $\dmax$. Assuming the thoretical prediction $\nu = 1.25$, the parameter $\alpha=2$ has the best agreement with the universal IR behavior. In Appendix \ref{sec:appendix-universal-IR} we discuss different choices of parameters and theoretical input $\nu$. If $\alpha$ is much above 2, then the IR scale does not look universal for different $\dmax$, though for values between 1 and 2 the scaling collapse is not much worse than at $\alpha=2$. This IR ansatz also has a preference for the theoretical critical exponent $\nu = 1.25$. Far away from $\nu = 1.25$ there is no $\alpha$ that can realize both (\ref{eq:universal-above-IR}) and (\ref{eq:universal-below-IR}).

The numerical result suggests that the IR scale is $\Lambda_{\rm IR} \sim \dmax^{-\alpha}$ for $\alpha$ near 2. Above the IR scale there may be other truncation effects that have not converged at $\dmax = 28$. Fig.\ \ref{fig:universalIR} tells us that the IR scale as well as the critical  exponent above the IR scale are clean and are ideal observables at finite truncation.

\section{Spontaneously Broken SUSY}
\label{sec:susy-breaking-phase}

In 
Fig.\ \ref{fig:spectrum}
we see the emergence of a massless phase as we dial $\bar g > \bar g_*$. 
There is a fuzzy region at the vicinity of the TIM critical point $\bar{g}_*$, where the mass gap is still fluctuating.
For greater $\bar g$, the gap obviously vanishes, and so has the level spacing.
The IR spectrum at large coupling nearly forms a continuum. 
From the discussion in Section \ref{sec:rg-flow} we expect the phase to be the SUSY-breaking $\mathbb{Z}_2$-preserving phase. The theory has massless goldstino, so the IR is in the same universality class as the 2D Ising theory.

\begin{figure}[h!] 
\centering\includegraphics[width=0.95\columnwidth]{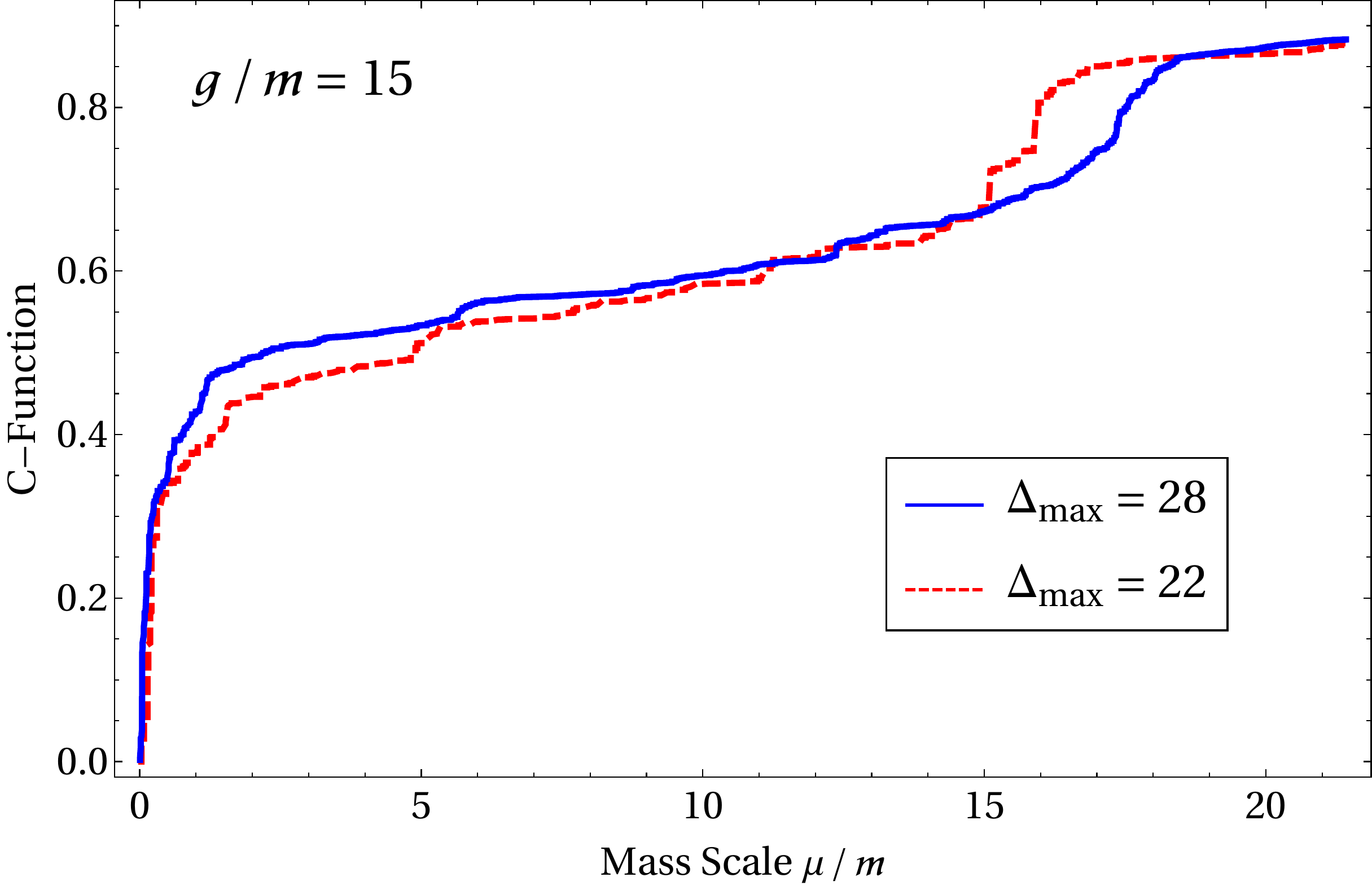}
\caption{\label{fig:ising-extrapolation}
The $C$-function of the SUSY broken phase at $\bar g = 15$. The coupling is chosen to be far beyond the critical point,
$\bar g \gg \bar g_*$, where $\bar g_* \approx 1.5$, so that the TIM critical behavior is not present. At this coupling, the spacing between individual mass eigenvalues is smaller than the resolution of this plot, so the plot shows continuous curves. The blue solid curve is the numerical $C$-function at the largest truncation level $\dmax=28$. The red dashed curve is computed at a lower truncation level $\dmax=22$. The difference between the two curves qualitatively reflects the convergence of the $C$-function at different mass scales $\mu$. At very large and very small $\mu$, the result is still sensitive to truncation. At intermediate $\mu$, the function stabilizes between $0.4 < C(\mu) < 0.6$, in agreement with the Ising model central charge $c_{\rm Ising}=0.5$. The leading deviation from the Ising prediction is due to the higher-dimensional deformation $T\bar{T}$, which can lift the asymptotic value of the $C$-function above $c_{\rm Ising}$ as $\dmax \rightarrow \infty$.
}
\end{figure} 

\begin{figure}[h!] 
\centering\includegraphics[width=0.95\columnwidth]{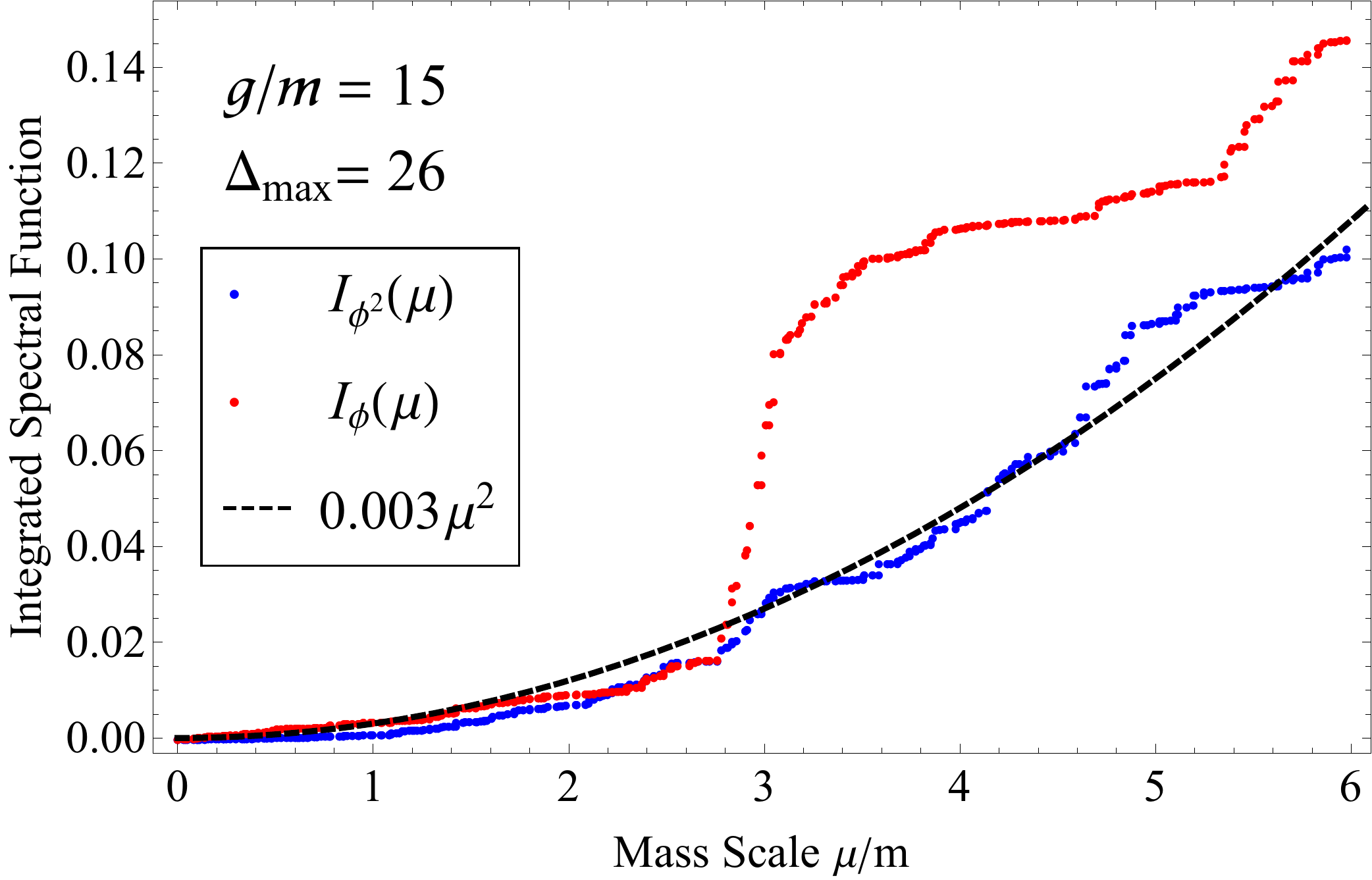}
\caption{\label{fig:ising-phiAndPhi2}
Integrated spectral functions for $\phi$ and $\phi^2$ at $\bar{g}=15$ and $\Dmax=26$.  Note that both follow the power-law behavior $\mu^2$ as expected from the fact that neither $\phi$ nor $\phi^2$ creates the $\sigma$ operator in the SGNY model, and therefore both operators flow to the $\epsilon$ operator from the Ising model at low energies.  The spectral function $I_\phi(\mu)$ has been rescaled to more easily compare its shape with that of $I_{\phi^2}(\mu)$.  This plot indicates that for $\mu \lesssim 3m$ at this value of $\bar{g}$, the theory has approached the Ising model.
}
\end{figure} 

We would like to study the spectral function and compare it to the Ising model in the IR. 
We first compute the $C$-function from the spectral function of $T_{--}$.  The $C$-function is a constant at the IR fixed point equal to the central charge of Ising $c_{\rm Ising} = \frac{1}{2}$. At higher mass scales  the $C$-function gets contributions from the UV deformation. In the $\mathbb{Z}_2$-preserving phase, the theory respects two $\mathbb{Z}_2$ symmetries. One is the spin $\mathbb{Z}_2$ symmetry, under which $\sigma$ is odd and $\epsilon$ is even. Focusing on the even sector, we have a second $\mathbb{Z}_2$ symmetry, the Kramers-Wannier duality, under which $\epsilon$ is odd. It is the second $\mathbb{Z}_2$ that protects the fermion mass on the $\bar g > \bar g_*$ side. The leading UV deformation that preserves both $\mathbb{Z}_2$ symmetries is the $T\bar T$ operator. The  $T\bar T$ deformation has a positive contribution to the $T_{--}$ spectral function.
The numerical result of the $C$-function is shown in Fig.\ \ref{fig:ising-extrapolation}.
At finite $\dmax$ the truncation effect shuts down the spectral function in the deep IR. We see that as $\dmax$ increases the IR region of the $C$-function flattens out and approaches $c_{\rm Ising} = \frac{1}{2}$. 

We study more operators in Fig.\ \ref{fig:ising-phiAndPhi2}. As is discussed in Section \ref{sec:rg-flow}, the spin $\mathbb{Z}_2$-odd operators cannot be constructed as local operators made from products of $\phi$s and $\psi$s in the UV. So, the $\sigma$ operator should never appear in the spectral function of such operators. In the IR fixed point, the spectral function of operators $\phi$ and $\phi^2$ should both be dominated by the most relevant operator $\epsilon$. From dimensional analysis $\braket{\epsilon\epsilon}\sim \mu^{2h_\epsilon}$, where  $h_\epsilon = \frac{1}{2}$, so the integrated spectral function should have the scaling behavior $I_{\epsilon}(\mu)\sim \mu^2$. Fig.\ \ref{fig:ising-phiAndPhi2} shows that $\phi$ and $\phi^2$ both match to this behavior at the lowest states.

In studying the spectral function in the SUSY-breaking phase we take $\bar g =15$, which is significantly larger than the critical coupling $\bar g_*$.
Unlike the TIM, where we have to tune $\bar g$ to the critical point, in the SUSY-breaking phase all RG flows with $\bar g > \bar g_*$
have the same IR fixed point. The finite $\dmax$ truncation introduces both a UV and an IR cutoff, so we have to deal with the fact that the numerical spectrum only has access to a finite range of the RG flow. Near the TIM critical point, it is likely that this range will be dominated by the TIM. Therefore, we move away from the TIM fixed point by taking $\bar g \gg \bar g_*$ in order to have better resolution at the Ising fixed point. 

The $C$-function is computed at $\dmax=28$, where the plot has stabilized to show the qualitative trend. 
If we would like to extrapolate the $C$-function to $\dmax \rightarrow \infty$, it will require a reliable model capturing corrections to both the strength and the position of each spectral line, which is beyond the scope of this paper. Instead, we choose to simply 
show two different $\dmax$ results to indicate how much we expect the function to change as we increase $\dmax$.

\section{Conclusion and Future Directions}

\begin{figure}[t!]
\centering
\includegraphics[width=0.95\columnwidth]{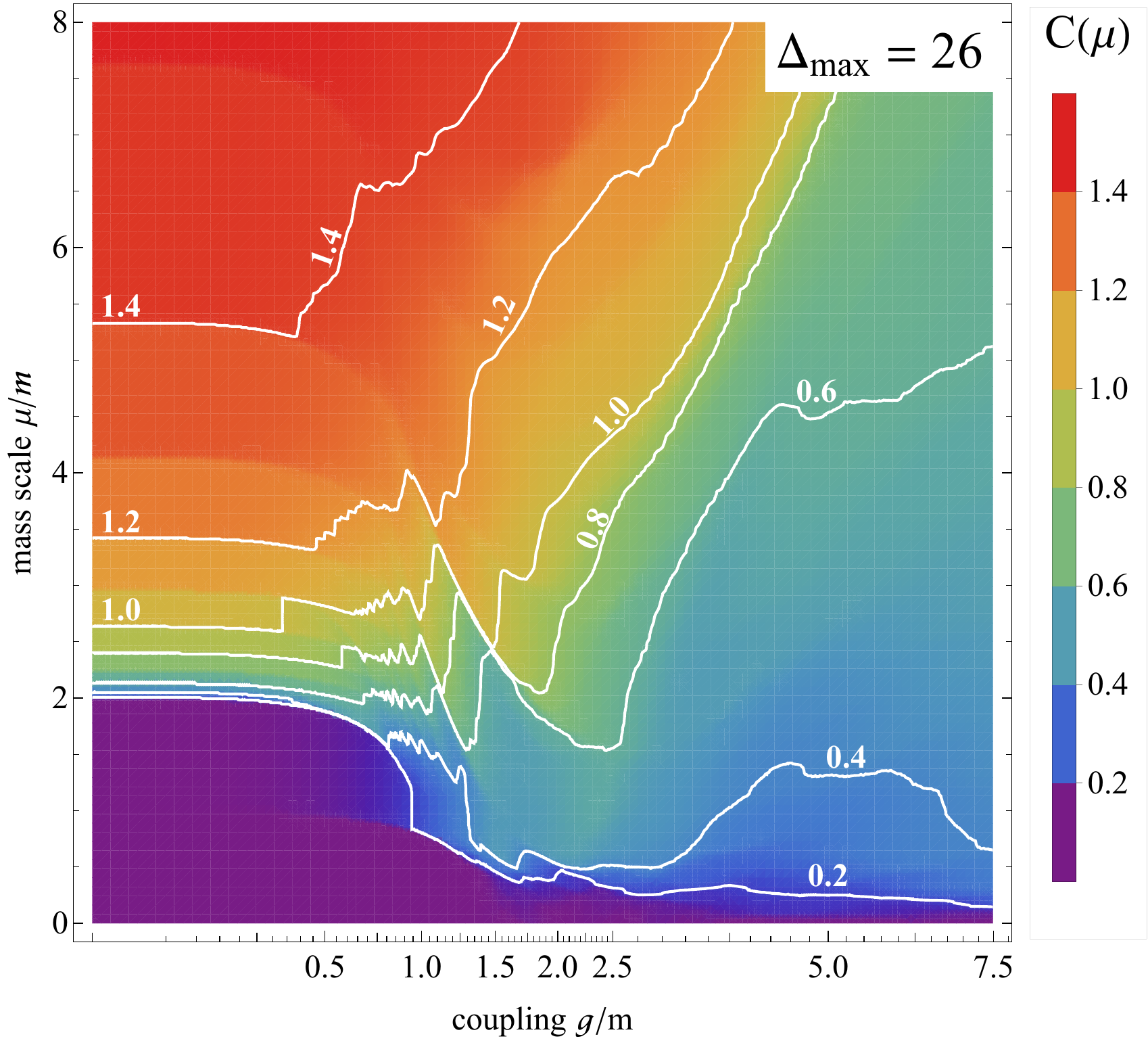}
\caption{\label{fig:heatPlot}
		Zamolodchikov $C$-function at different couplings. 
}
\end{figure}

Our analysis of the SGNY model represents the first application of LCT to a theory where the UV CFT is supersymmetric, and in fact is the first application with both fermions and scalars in the UV Lagrangian.  Our numeric results pass many nontrivial checks by comparing to analytic results in different regimes of the theory.  One of the most interesting observables we compute is the $C$-function, shown as a function of coupling $\bar{g}$ and scale $\mu$ in Fig.\ \ref{fig:heatPlot}.  We have seen in Sections~\ref{sec:massive} and \ref{sec:susy-breaking-phase} various slices of this plot at particular values of $\bar{g}$ and confirmed its level of convergence. The color scheme of this plot was chosen to match that of Fig.\ \ref{fig:PhaseDiagramCartoon}, allowing us to clearly see how our numerical results confirm the conceptual picture presented in Section~\ref{sec:rg-flow}. 

It is remarkable to us that the formulation used here, where we construct the Hamiltonian by squaring one of the supercharges computed in a truncated basis, works on both sides of the phase transition; in lightcone quantization, correctly dealing with changes in the vacuum from one minimum of the potential to another can be quite subtle. It would be useful to understand whether this is a general feature of SUSY theories, especially in higher dimensions, such that we can use supercharges to study a wide range of phase transitions. One further advantage of using $Q_+$ to construct the Hamiltonian is that it allows us to avoid integrating out $\chi$, thereby keeping all of the interactions local.

Relatedly, we would like to study the far side of the phase transition, where SUSY is spontaneously broken, in more detail. Near the critical point, the IR behavior should match the integrable flow from TIM to the Ising model, for which many correlation functions have been computed analytically~\cite{Delfino:1994ea,Fioravanti:2000xz,Horvath:2016nbm}. Even far away from the critical point, however, the IR behavior should be accurately modeled by a $T\bar{T}$ deformation of the Ising model, and it would be interesting to precisely match this effective description to spectral functions. In addition, one could explicitly break SUSY with a $\phi^2$ deformation, which in the IR should be equivalent to deforming the Ising model by $\epsilon$.

The numerical results in this paper relied crucially on several technical innovations for constructing the conformal truncation basis and computing the Hamiltonian matrix elements, which will be described in detail in \cite{pedagogical}.  The methods used in a previous paper \cite{Phi4Paper}, running for approximately one day, would allow only about $7000$ basis states; the highest truncation level in this work is $\dmax=28$, which includes $69568$ basis states.  The matrix elements of $Q_+$ 
are computed in series and the matrix diagonalization is parallelized.
 The matrix $Q_+$ has approximately $2\times 10^8$ nonzero elements. Generating the basis and the matrix elements, which is required only once, takes one day on a desktop. For each coupling value $\bar g$, exactly diagonalizing the $\dmax=28$ $Q_+$ matrix to obtain the full set of eigenvalues and eigenvectors takes two hours on a 28-core cluster. 
There are a number of additional techniques that could be implemented in the future to improve the quality of the results from this kind of analysis.  Some improvements could come simply from increased computational power, for instance by further parallelizations of the code.  Moreover, all the numeric computations in this paper were done in Mathematica and could potentially be sped up by moving to a more efficient programming language.

Beyond the above, there are more conceptual improvements that could be made to the convergence of LCT. For instance, one could potentially adapt the renormalization techniques of~\cite{Feverati:2006ni,Watts:2011cr,Giokas:2011ix,Lencses:2014tba,Hogervorst:2014rta,Rychkov:2014eea,Rychkov:2015vap,Elias-Miro:2015bqk,Elias-Miro:2017tup,Elias-Miro:2017xxf,Rutter:2018aog,Hogervorst:2018otc} to modify the truncated Hamiltonian to include the effects of operators above $\Dmax$. In this work, we observed the emergence of a universal IR scale due to truncation. It would be interesting to better understand the effects of this IR scale, in order to extrapolate results in $\Dmax$.

\appendix

\section*{Acknowledgments}
We thank Nikhil Anand, Luca D\'elacretaz, Kyriakos Grammatikos, Matthijs Hogervorst, Jared Kaplan, Zuhair Khandker, 
Jo\~{a}o Penedones, and Yijian Zou for useful conversations. ALF, EK, and YX were supported in part by the US Department of Energy Office of Science under Award Number DE-SC0015845, and ALF in part by a Sloan Foundation fellowship.  MW is partly supported by the National Centre of Competence in Research SwissMAP funded by the Swiss National Science Foundation. The authors were also supported in part by the Simons Collaboration Grant on the Non-Perturbative Bootstrap.

\onecolumngrid
 
\section{Free and Perturbative Theory Details}
\label{app:FreeAndPert}


Here we provide some details of the LCT calculations done for small ($n\le 2$) particle number in the free theory and in perturbation theory.  First, we quote the result for the wavefunctions for two-particle $[\phi \phi]$, $[\psi \psi]$, and $[\phi \psi]$ primary operators.
Particle states satisfy 
\be
|p_1 , \dots , p_n \> = \sqrt{2 p_1 \dots 2 p_n} a_{p_1}^\dagger \dots a_{p_n}^\dagger |0 \>.
\ee
with $\< p | p'\> = (2p)(2\pi) \delta(p-p')$.  The momentum space wavefunctions $f(p_1, \dots, p_n)$ for an operator $\CO$ are defined by
\be
\< p_1, \dots , p_n| \CO(0)\> \equiv p_1 \dots p_n f_{\CO}(p_1, \dots, p_n).
\ee
For two-particle states, they are simply Jacobi polynomials $P_k^{(\alpha, \beta)}(x)$:
\bea
\left[\phi \phi\right]_k &:& f^{(\phi \phi)}_k(p_1) =  \widehat{P}_{k}^{\left(1,1\right)}\left(1-2p_1\right) , \qquad k = 0, 2, 4, \dots \nn\\
\left[\psi \psi\right]_k &:& f^{(\psi \psi)}_k(p_1) = \widehat{P}_{k}^{\left(2,2\right)}\left(1-2p_1\right) , \qquad k = 1, 3, 5, \dots \nn\\
\left[\phi \psi\right]_k &:& f^{(\phi \psi)}_k(p_1) = \widehat{P}_{k}^{\left(1,2\right)}\left(1-2p_1\right) , \qquad k = 0, 1, 2, \dots .
\eea
where we have eliminated $p_2$ by using $p_1 +p_2=p=1$ in our momentum frame,  and
\bea
&& \widehat{P}_{k}^{\left(\alpha,\beta\right)}\left(x\right) \equiv \mu_k^{(\alpha,\beta)} P_{k}^{\left(\alpha,\beta\right)}\left(x\right), \\
&& \mu_k^{(\alpha,\beta)} \equiv  \frac{1}{\sqrt{\cal N}} \sqrt{\frac{\Gamma\left(k+1\right)\Gamma\left(k+\alpha+\beta+1\right)\Gamma\left(2k+\alpha+\beta+2\right)}{\Gamma\left(k+\alpha+1\right)\Gamma\left(k+\beta+1\right)\Gamma\left(2k+\alpha+\beta+1\right)}}. 
\eea
Here, ${\cal N}$ is a constant prefactor that can be absorbed into the normalization of the fields. Setting it to ${\cal N}=1$, the states are normalized:
\bea
\frac{\< [\phi \phi]_k | [\phi \phi]_{k'}\>}{2p(2\pi) \delta(p-p')} &=&  \int_0^1 dx x (1-x) f^{(\phi \phi)}_k(x) f^{(\phi \phi)}_{k'}(x) = \delta_{k k'} , \nn\\
\frac{\< [\psi \psi]_k | [\psi \psi]_{k'}\>}{2p(2\pi) \delta(p-p')} &=&  \int_0^1 dx x^2 (1-x)^2 f^{(\psi \psi)}_k(x) f^{(\psi \psi)}_{k'}(x) = \delta_{k k'} , \nn\\
\frac{\< [\phi \psi]_k | [\phi \psi]_{k'}\>}{2p(2\pi) \delta(p-p')} &=&  \int_0^1 dx x (1-x)^2 f^{(\phi \psi)}_k(x) f^{(\phi \psi)}_{k'}(x) = \delta_{k k'} , \nn\\
\eea

Matrix elements, exact spectrum and $\Delta_{\rm max}^{-2}$ IR scale, spectral functions.

The mass term matrix elements of $2 P_- P_+$ in the two-particle sector are 
\bea
(2 P_+)^{(\phi \phi)}_{k k'} &=&  \int_0^1 dx  f^{(\phi \phi)}_k(x) f^{(\phi \phi)}_{k'}(x) =  2 m^2 \sqrt{\frac{(k_m+1)(k_m+2)}{(k_M+1)(k_M+2)} (2k+3)(2k'+3)} ,  \nn\\
(2 P_+)^{(\phi \psi)}_{k k'} &=& \int_0^1 dx (1-x)  f^{(\phi \psi)}_k(x) f^{(\phi \psi)}_{k'}(x) = m^2\left(2 +  (-1)^{k+k'}\left(\frac{k_m+2}{k_M+2}\right) \right)\left(  \sqrt{ \frac{(k_m+1)(k_m+3)}{(k_M+1)(k_M+3)} (k+2)(k+2)}\right) , \nn\\
  (2 P_+)^{(\psi \psi)}_{k k'} &=&\int_0^1 dx x(1-x)f^{(\psi \psi)}_k(x) f^{(\psi \psi)}_{k'}(x)=  m^2\sqrt{ \frac{ (k_m+1)_4}{(k_M+1)_4} (5+2k)(5+2k')} ,
  \label{eq:TwoPclMassMat}
  \eea
where $k_m= {\rm min}(k,k')$ and $k_M = {\max}(k,k')$.


 The characteristic polynomial of the two-fermion mass matrix $M$ truncated at $k \le \Delta_{\rm max}$ is
\be
\det(M- m^2 x) = (-x)^{\frac{\Delta_{\rm max}+1}{2}} {}_2F_1\left( - \frac{\Delta_{\rm max}+1}{2}, 3+ \frac{\Delta_{\rm max}}{2}, 2 , \frac{4}{x} \right) = \frac{2  (-x)^{\frac{\Delta_{\rm max}+1}{2}}P_{\frac{\Delta_{\rm max}+1}{2}}^{(1,\frac{1}{2})}(1-\frac{8}{x})}{\Delta_{\rm max}+3}  \nn\\
\ee
and the mass-squared spectrum $m_n^2 = m^2 x_n$ is given by the zeros $x_n$ of this polynomial. We can obtain the spectrum at large $\Delta_{\rm max}$ through a useful asymptotic relation for the Jacobi polynomials:
\be
P_n^{(\alpha, \beta)}(\cos \theta) = \frac{\cos \left\{ \left[ n + \frac{1}{2} (\alpha + \beta + 1) \right] \theta - \left( \frac{1}{2} \alpha + \frac{1}{4} \right) \pi \right\} }{\sqrt{ \pi n} \left( \sin \frac{\theta}{2} \right)^{\alpha + \frac{1}{2}} \left( \cos \frac{\theta}{2} \right)^{\beta + \frac{1}{2}}} + {\cal O}(n^{-\frac{3}{2}})
\ee
It is easy to solve for the values of $\theta$ for which the leading term at large $n$ vanishes.  Applied to the case at hand, 
 at large $\Delta_{\rm max}$ we see that the eigenvalues are approximately
\be
m_n^2 = m^2 x_n = 4 m^2 \sec^2 \left( \frac{2 \pi n}{2 \Delta_{\rm max}+7} \right).
\ee

For reference, we also provide the following expressions for the overlaps for some states and operators discussed in the main body of the paper.  First, the overlap between the operator $\psi \chi$ and the two-particle states $[\psi \psi]_k$ in the truncation basis is
\bea
\< \psi \chi | [ \psi \psi]_k\> 
 &=& \int_0^1 dx x(1-x) \left( \frac{1}{x} - \frac{1}{1-x}\right) f_k(x) = 2 \sqrt{\frac{(k+1) (k+4) (2 k+5)}{(k+2) (k+3)}} .
 \label{eq:psichiMatEl}
\eea
This overlap enters in the calculation of the spectral function $\rho_{\psi \chi}$ for $\psi \chi$ in the free theory.
Second, the matrix element of the interaction $ \phi \psi \frac{1}{\partial} \psi$ between a single-fermion state and a two-partile $[\phi \psi]_k$ state in the truncation basis is
\bea
\< \psi | \phi \psi \frac{1}{\partial} \psi | [ \phi \psi]_k \> &=& \int_0^1 dx  (1-x) \left( 1+ \frac{1}{1-x} \right) f_k(x) =  (-1)^k \sqrt{\frac{2}{(k+1) (k+2) (k+3)}} \left(k+2+(-1)^k\right)^2 .
\eea
This matrix element enters into the calculation of the divergence in the shift in the $\psi$ mass at second order in perturbation theory.

\section{TIM Form Factors}
\label{app:TIMFormFactors}

The phase transition between the $\mathbb{Z}_2$-breaking and SUSY-breaking phases of the SGNY model is in the same universality class as the tricritical Ising model. In the vicinity of this critical point, we thus expect the IR regime to correspond to a deformation of TIM by its only relevant, SUSY-invariant scalar: the vacancy operator $\epsilon'$,
\be
S_\IR \approx S_{\textrm{TIM}} + \lambda \int d^2x \, \epsilon'(x).
\ee
This particular deformation of TIM is integrable, and falls into a general class of integrable deformations of minimal models by $\Phi_{1,3}$ Virasoro primaries~\cite{Zamolodchikov:1987ti,Kastor:1988ef}. For $\lambda > 0$, the RG flow preserves the spin-reversal $\mathbb{Z}_2$, but SUSY is spontaneously broken, resulting in a massless theory described by the 2D Ising model in the IR~\cite{Zamolodchikov:1991vx}. For $\lambda < 0$, however, SUSY is preserved, but now $\mathbb{Z}_2$ is spontaneously broken, with three degenerate ground states. The spectrum of this theory contains massive ``kinks'' connecting the different ground states, whose S-matrix is described by the restricted solid-on-solid (RSOS) scattering theory~\cite{Zamolodchikov:1991vh}. These kinks do not form bound states, so in the sector with periodic boundary conditions, the lowest states in the spectrum correspond to the continuum of unbound kink-antikink states.\footnote{This spectrum was studied numerically in~\cite{Lassig:1990xy,Lepori:2008et}.}

The spectral functions of local operators in the deformed theory can be expressed in terms of form factors corresponding to the overlap of these operators with multi-kink states,
\be
F^{(n)}_\Ocal(\theta_1,\ldots,\theta_n) \equiv \<\Omega|\Ocal(0)|\theta_1,\ldots,\theta_n\>,
\ee
where $\theta_i$ is the rapidity of a single (anti)kink. For a given operator $\Ocal$, the spectral function can thus be computed by summing over all possible intermediate multi-kink states,
\be
\rho_\Ocal(\mu) = \sum_n \fr{1}{n!} \int \fr{d\theta_1 \cdots d\theta_n}{(2\pi)^n} (2\pi)^2 \de^2(P-p_1-\cdots-p_n) |F^{(n)}_\Ocal(\theta_1,\ldots,\theta_n)|^2.
\ee

This sum over multi-kink states converges very rapidly, so in practice we can accurately approximate the spectral function by only including the first contribution in this sum (this was demonstrated explicitly for the case of TIM in~\cite{Delfino:1999et}). For local operators such as the stress-energy tensor, the first contribution to these spectral functions comes from kink-antikink states ($n=2$). The associated form factor is only a function of the difference in rapidities, which we can write in terms of the kink mass $\mkink \equiv \fr{\mgap}{2}$ and total invariant mass $\mu$ as
\be
|\theta_1 - \theta_2| = 2\log\bigg(\fr{\mu+\sqrt{\mu^2 - 4\mkink^2}}{2\mkink}\bigg).
\ee
The resulting approximate spectral function (where we neglect contributions from higher-kink states) thus takes the form
\be
\rho_\Ocal(\mu) \approx \fr{\bigg|F^{(2)}_\Ocal\Big(2\log\fr{\mu+\sqrt{\mu^2-4\mkink^2}}{2\mkink}\Big)\bigg|^2}{\pi\mu\sqrt{\mu^2-4\mkink^2}}.
\ee
We just need to compute the kink-antikink form factor $F^{(2)}_\Ocal(\theta)$, which can be fixed by analyticity, unitarity, and crossing symmetry (see~\cite{Mussardo} for a thorough review of such methods).

For example, we can consider the trace of the stress-energy tensor, $T_{+-} \sim \mkink^{\fr{4}{5}} \epsilon'$. This operator has the resulting kink-antikink form factor~\cite{Delfino:1999et}
\be
F^{(2)}_{T_{+-}}(\theta) = -\fr{\mkink^2}{2^{\fr{11}{4}}} e^{\fr{i\theta\log2}{4\pi}} \fr{\sinh\theta}{\sinh \fr{1}{4}(\theta-i\pi)} \exp[-A(\theta)],
\ee
where $A(\theta)$ is defined as the integral
\be
A(\theta) \equiv \int_0^\infty \fr{dx}{x} \fr{\sinh\fr{3x}{2}}{\sinh2x\cosh\fr{x}{2}} \fr{\sin^2\fr{(i\pi-\theta)x}{2\pi}}{\sinh x}.
\ee
In practice, this integral can be evaluated numerically for a range of values of $\theta$, in order to obtain the resulting spectral function.

This kink-antikink form factor was used to compute the TIM prediction for the integrated spectral function of $T_{+-}$ in figure~\ref{fig:TraceGrid} (black dashed line). Via the Ward identity, we also used this form factor to compute the TIM prediction for the Zamolodchikov $C$-function
\be
C(\mu) = \fr{12\pi}{P_-^4} \int d\mu'^2 \rho_{T_{--}}(\mu') = 48\pi \int \fr{d\mu'^2}{\mu'^4} \rho_{T_{+-}}(\mu'),
\ee
shown in figure~\ref{fig:CFuncGrid} (black dashed line).

While we have technically only included the leading contribution to the spectral function for these TIM predictions, we can check the validity of this approximation by looking at the asymptotic behavior of $C(\mu)$ in the UV, finding
\be
C(\mu) \approx 0.69 \quad (\mu \ra \infty),
\ee
compared with the exact value of $c_\TIM = 0.7$. We thus see that all higher-kink contributions provide at most a percent-level correction to the TIM spectral functions.


\section{Doubly-Integrated Spectral Functions}
\label{app:doubleInts}

\begin{figure}[t!]
\begin{center}
\includegraphics[width=0.45\textwidth]{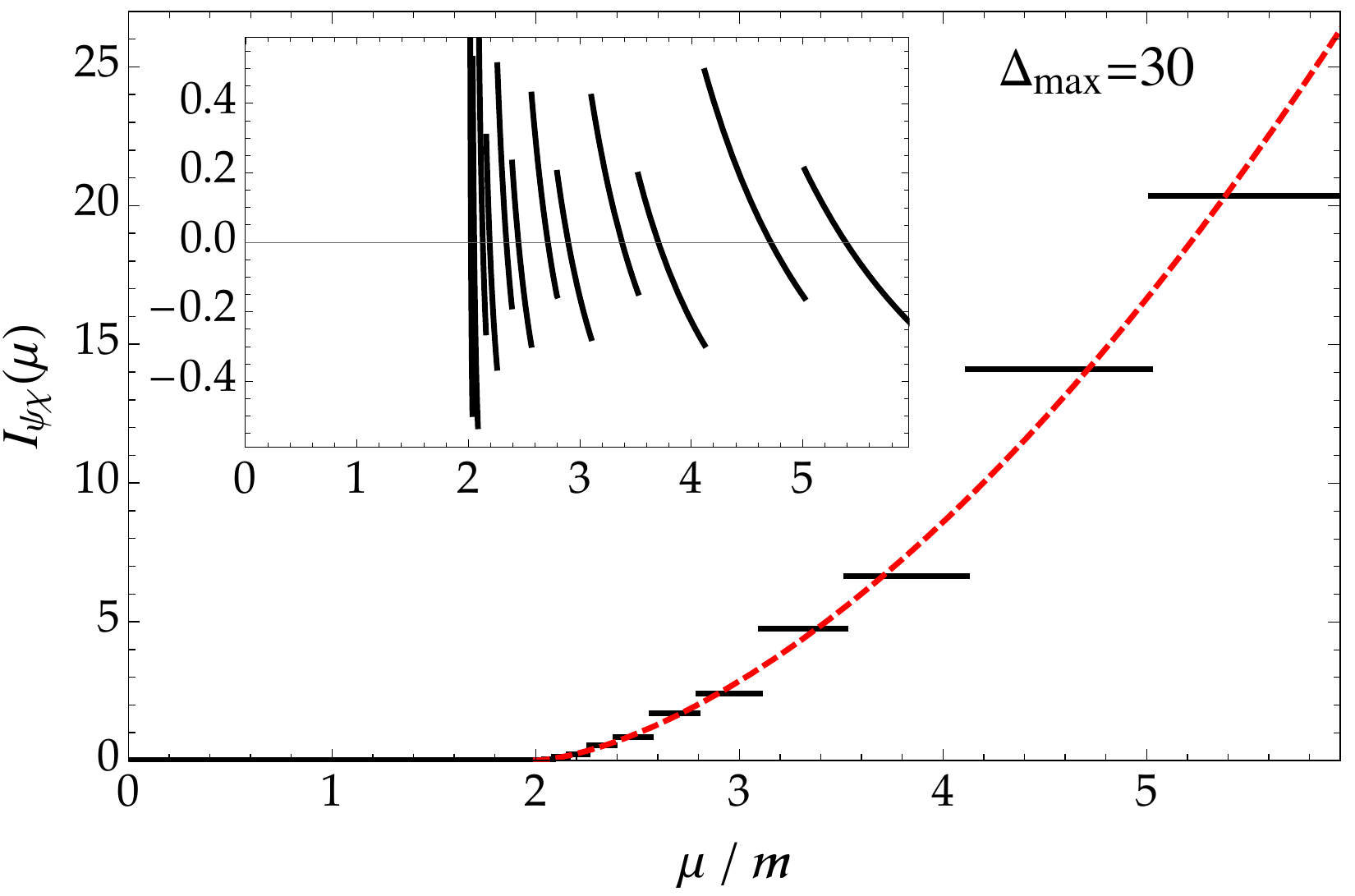}~~
\includegraphics[width=0.45\textwidth]{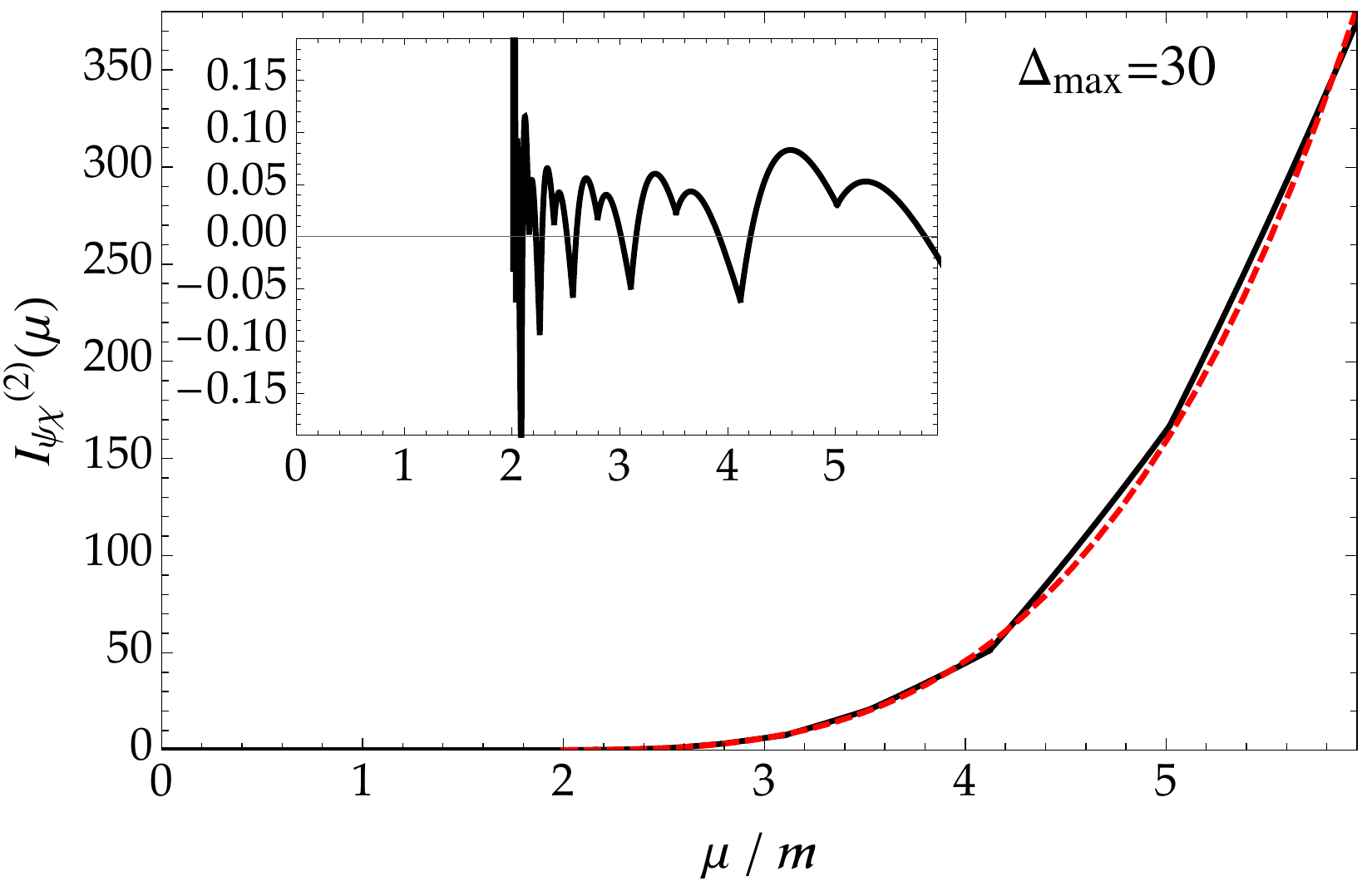}
\caption{Integrated spectral function $I(\mu)$ and doubly-integrated spectral function $I^{(2)}(\mu)$ for the operator $\psi \chi$ in the free massive theory, where the spectral function in the continuum is $\rho_{\psi \chi}(\mu)\propto (1-4m^2/\mu^2)^{1/2}$.  The truncation result at $\Delta_{\rm max}=30$ is the black solid line, and the continuum result is shown in red dashed line.  The relative error $\delta I \equiv I_{\rm trunc}/I_{\rm analytic}-1$ is shown in the inset.}
\label{fig:SpecFuncPsiChi}
\end{center}
\end{figure}

\begin{figure*}[t!]
\centering
\includegraphics[width=0.85\textwidth]{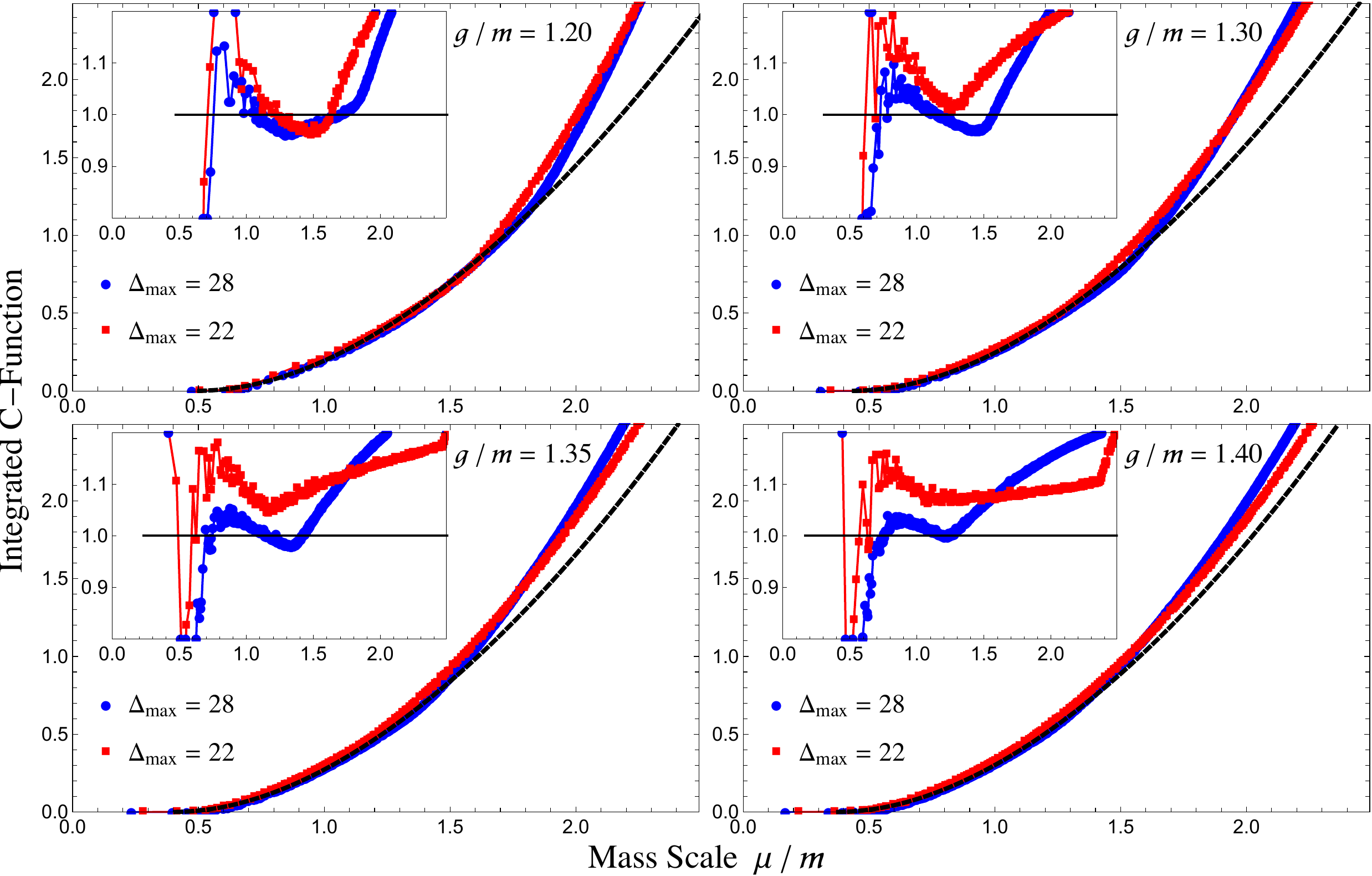}
\caption{\label{fig:CFunc2Grid} 
Integrated $C$-function, as defined in eq.~\eqref{eq:IntCFunc}. Main plots: integrals of the truncation data and theoretical TIM prediction from figure~\ref{fig:CFuncGrid}. Insets: ratio of the truncation data at $\Dmax=28$ (blue dots) and $\Dmax=22$ (red squares) to the IR theoretical prediction (black line). The values of $\mgap$ for the theoretical prediction were determined by fitting to the data.
}
\end{figure*} 

\begin{figure*}[t!]
\centering
\includegraphics[width=0.85\textwidth]{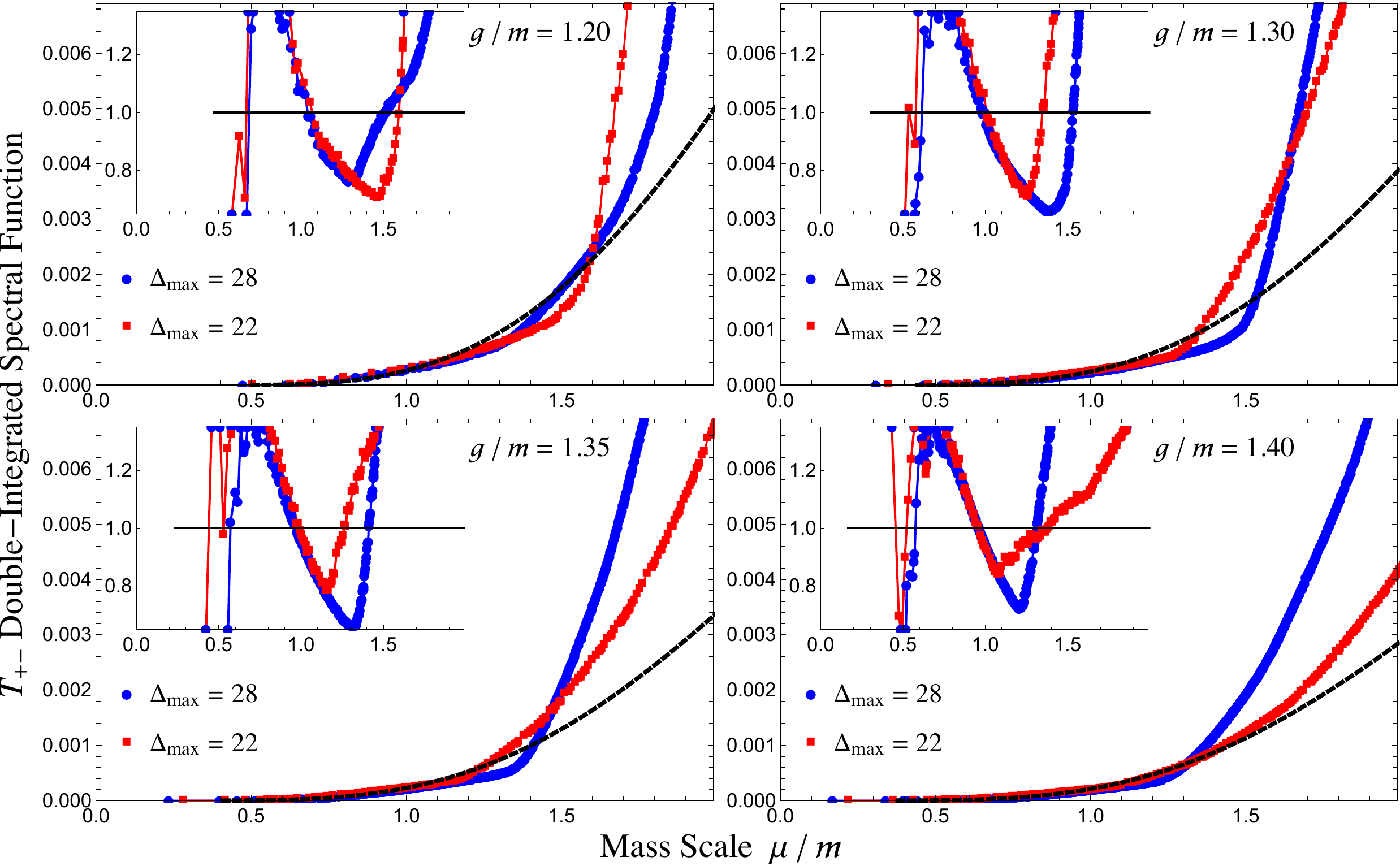}
\caption{\label{fig:Trace2Grid} 
Double-integrated spectral function for $T_{+-}$, as defined in eq.~\eqref{eq:DoubleIntT+-}. Main plots: integrals of the truncation data and theoretical TIM prediction from figure~\ref{fig:TraceGrid}. Insets: ratio of the truncation data at $\Dmax=28$ (blue dots) and $\Dmax=22$ (red squares) to the IR theoretical prediction (black line).
}
\end{figure*}

The spectral functions computed in a truncation framework generally have unphysical discontinuities due to the fact that often they are trying to approximate continuous spectra with discrete ones.  This fact is most apparent in the spectral functions themselves, which strictly speaking are sums over isolated delta functions for any finite truncation despite the fact that in the continuum limit most of these delta functions merge to form continuous functions.  Integrating at least one time is necessary in order to even plot the truncated spectral functions.  Integrating additional times has the advantage that not only do the resulting spectral functions become more smooth, but they generally also have reduced relative errors compared to the multiply-integrated spectral functions of the continuum theory.  

As a simple example of this feature, consider the spectral function for the operator $\psi \chi$ in the free massive theory.  The analytic result is eq. (\ref{eq:PsiChiLoop}).  To compute the result in truncation, we find the eigenvalues of the mass matrix $(2 P_+)^{(\psi \psi)}$ for two particle states, given in eq. (\ref{eq:TwoPclMassMat}), and  the matrix elements and spectrum for two-particle states,  given in (\ref{eq:psichiMatEl}), and compute the spectral function according to the general spectral function formula (\ref{eq:GenSpecFunc}). The integrated spectral function and doubly-integrated spectral functions,
\be
I_{\psi\chi}(\mu) \equiv \int_0^{\mu^2} d\mu'^2 \rho_{\psi \chi}(\mu'), \qquad I_{\psi\chi}^{(2)}(\mu) \equiv \int_0^{\mu^2} d\mu'^2 I_{\psi\chi}(\mu'),
\ee
respectively, are shown in Fig.\ \ref{fig:SpecFuncPsiChi}, and compared to the analytic results. The relative error of the doubly-integrated spectral function is significantly reduced compared to the integrated spectral function.

Next, we show similar additional integrations for some spectral functions in the interacting theory. In Fig.\ \ref{fig:CFuncGrid}, we showed the $C$-function, which is itself an integral of a spectral function.  We can perform an additional integration to obtain the function
\be
C^{(2)}(\mu) \equiv \int_0^{\mu^2} d\mu'^2 \, C(\mu').
\label{eq:IntCFunc}
\ee
Figure~\ref{fig:CFunc2Grid} shows the truncation results for $C^{(2)}(\mu)$, again compared to the TIM prediction. As we can see, these results are much smoother, and agree with the theoretical prediction to within a few percent, as we can see from the ratio shown in each inset. It is worth reiterating that the results in figure~\ref{fig:CFunc2Grid} are simply the integral of the results in figure~\ref{fig:CFuncGrid}. While taking this integral adds no new information, it allows us to see more clearly how well our truncation results match the TIM prediction at low energies. We can also easily see the scale at which the SGNY results deviate from the TIM description in the UV.

Similarly, we can smooth out the truncation data somewhat for the spectral function of the trace $T_{+-}$ by integrating a second time, to compute
\be
I^{(2)}_{T_{+-}}(\mu) \equiv \int_0^{\mu^2} d\mu'^2 \, I_{T_{+-}}(\mu').
\label{eq:DoubleIntT+-}
\ee
Figure~\ref{fig:Trace2Grid} shows this double-integrated spectral function, again compared to the TIM prediction. The ratio of the truncation data to the theoretical curve is shown in the insets, where we can see that the values agree to within roughly $25\%$. The reason this error is much larger than in figure~\ref{fig:CFunc2Grid} is that $T_{+-}$ is going to zero, so the numerical value for its spectral function is orders of magnitude smaller than that of $T_{--}$.

\section{Convergence of Mass Eigenvalues}

\begin{figure}[t!]
\centering
\includegraphics[width=0.45\columnwidth]{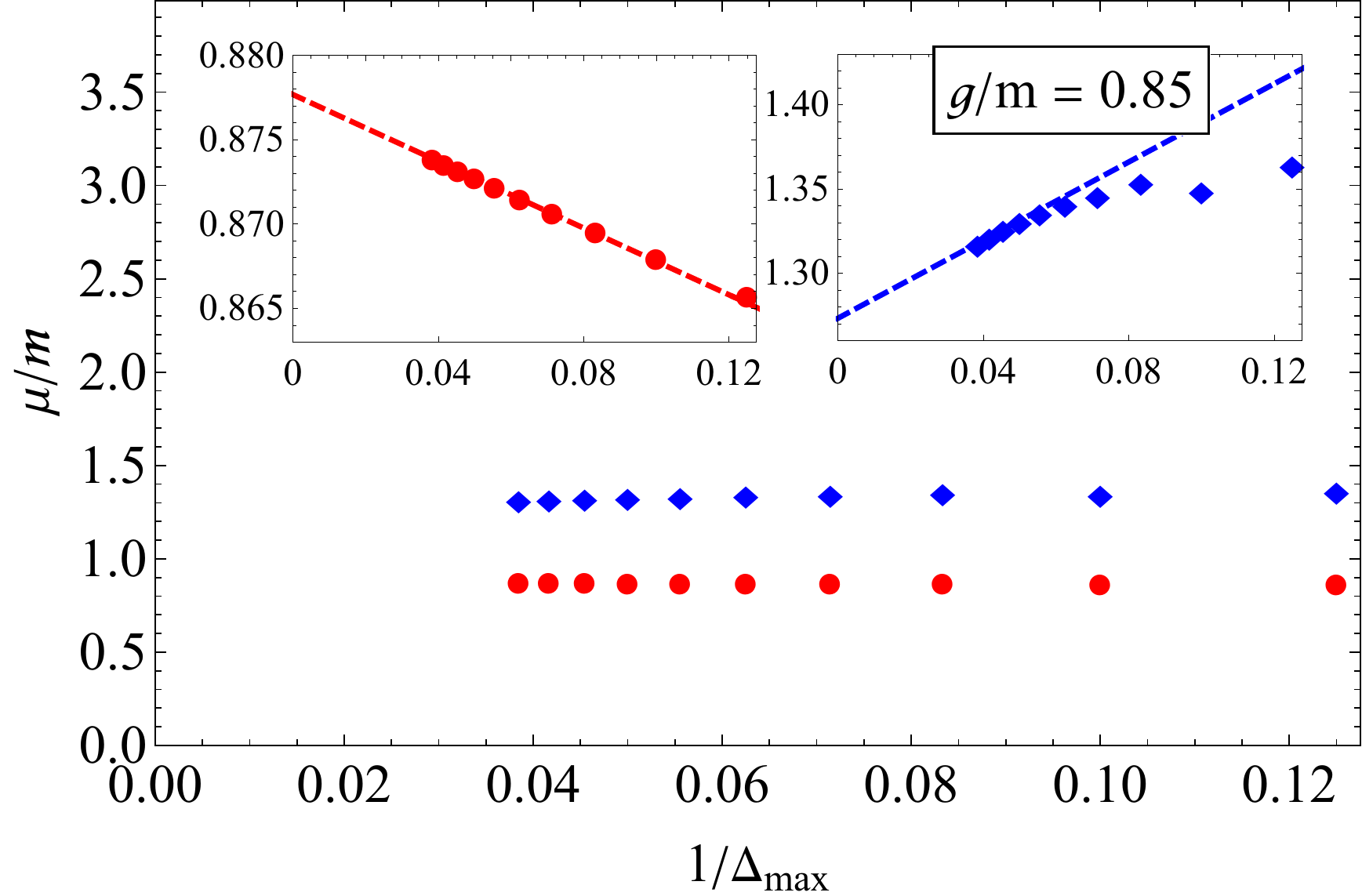} 
~~
\includegraphics[width=0.45\columnwidth]{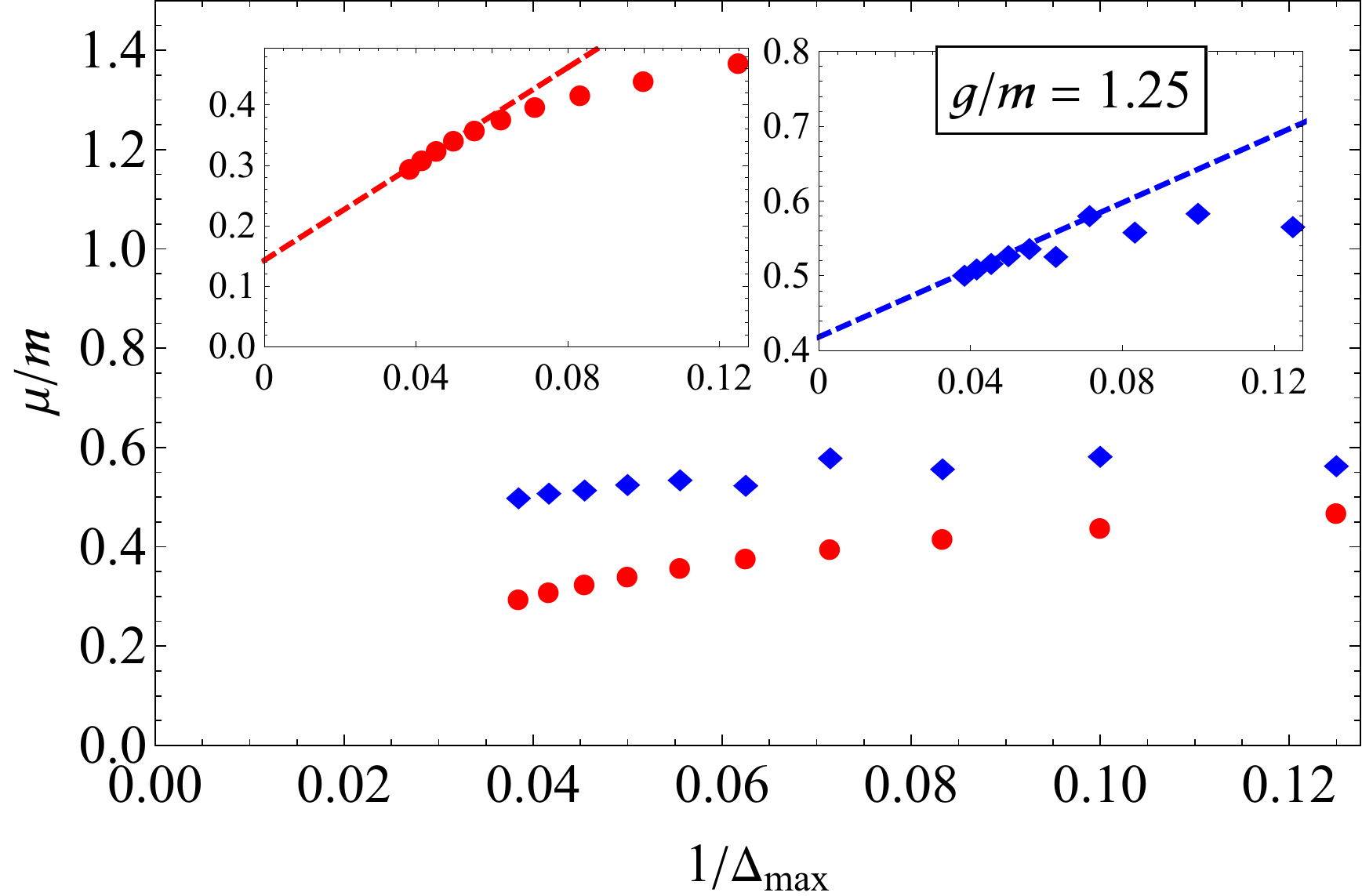}
\caption{\label{fig:convergence}
The convergence of the first (red circle) and second (blue diamond) lowest mass eigenvalues at strong couplings $\bar g = 0.85$ (left plot) and $\bar g = 1.25$ (right plot). The plot shows the eigenvalues at fixed $\bar g$ as a function of $1/\dmax$. The insets of the plots zoom in to each individual eigenvalue to show the dependence on $1/\dmax$. The dashed trend lines are linear fits of the last 4 points.
At the smaller coupling, $\bar g = 0.85$, both eigenvalues have converged well.  At the larger coupling, $\bar g = 1.25$, which is close to the TIM critical point, both eigenvalues are still changing as a function of $\dmax$. 
The lowest eigenvalue at $\bar g = 0.85$ has a clear linear dependence on $1/\dmax$. The others have slower convergence rates. The changing slope may suggest that the dependence on $1/\dmax$ has a different power, or the dependence is linear but the slopes have not converged to constants.
}
\end{figure}

In Section \ref{sec:mass-spectrum-interacting} we briefly discussed the convergence of the mass gap. We treated the mass gap at different values of $\bar g$ collectively and measured the change of the curve $\mgap(\bar g)$ due to $\dmax$. This strategy is useful in extracting the critical exponent. We would like to also study the convergence of individual mass eigenvalues.  

We study each mass eigenvalue as a function of $\dmax$  at fixed $\bar g$. At $\dmax\rightarrow\infty$, the eigenvalues should converge to constants. For sufficiently large $\dmax$, the shift due to truncation should fall off as a power law of $1/\dmax$. Based on the behavior of Fig.\ \ref{fig:spectrum} and Fig.\ \ref{fig:gap} we expect 
\begin{itemize}
	\item The convergence at small coupling $\bar g$ is better than at large $\bar g$. For each eigenvalue, its convergence is better before the collision with the continuum than after the collision. 
	\item As $\bar g$ increases, the continuum states start colliding with the lower spectrum from top to bottom. Thus the convergence is expected to get worse from top to bottom.
\end{itemize}
In Fig.\ \ref{fig:convergence} we take the lowest two mass eigenvalues (the red and blue curves in Fig.\ \ref{fig:spectrum}) as an example of the spectrum. We pick $\bar g=0.85$ as an example of the regime where the second state has merged into  the continuum and the first state has not collided yet, and we pick $\bar g=1.25$ as an example of the regime where both states have merged into the continuum. 
We plot each mass eigenvalue as a function of $1/\dmax$. The result matches our expectation.  The shift due to truncation is smaller at $\bar g = 0.85$ than at  $\bar g = 1.25$. In addition, the shift of the first mass eigenvalue at $\bar g = 0.85$ fits to a linear law of $1/\dmax$. In all of the other 3 cases, the convergence are worse. It is unclear from the data whether the fall-off of the truncation effect for the states in the continuum obeys a different power law, or $\dmax$ is not sufficiently large for the power law to dominate. 

Given this result one may be surprised why a straightforward fit to the critical exponent at each $\dmax$ works, even before the mass gap has converged. If the truncation effects between different $\bar g$ were uncorrelated, we would have no choice but to wait until $\mgap(\bar g)$ at all $\bar g$ have converged. 
In fact, it is likely that the truncation effects modify the physical observables in a universal manner that depends only on a mass scale set by $\dmax$. We discuss this in detail in
Section \ref{sec:universal-IR}.

\section{Parameter Dependence of the IR Scale}
\label{sec:appendix-universal-IR}

\begin{figure}[t!]
{
\centering
\includegraphics[width=0.95\columnwidth]{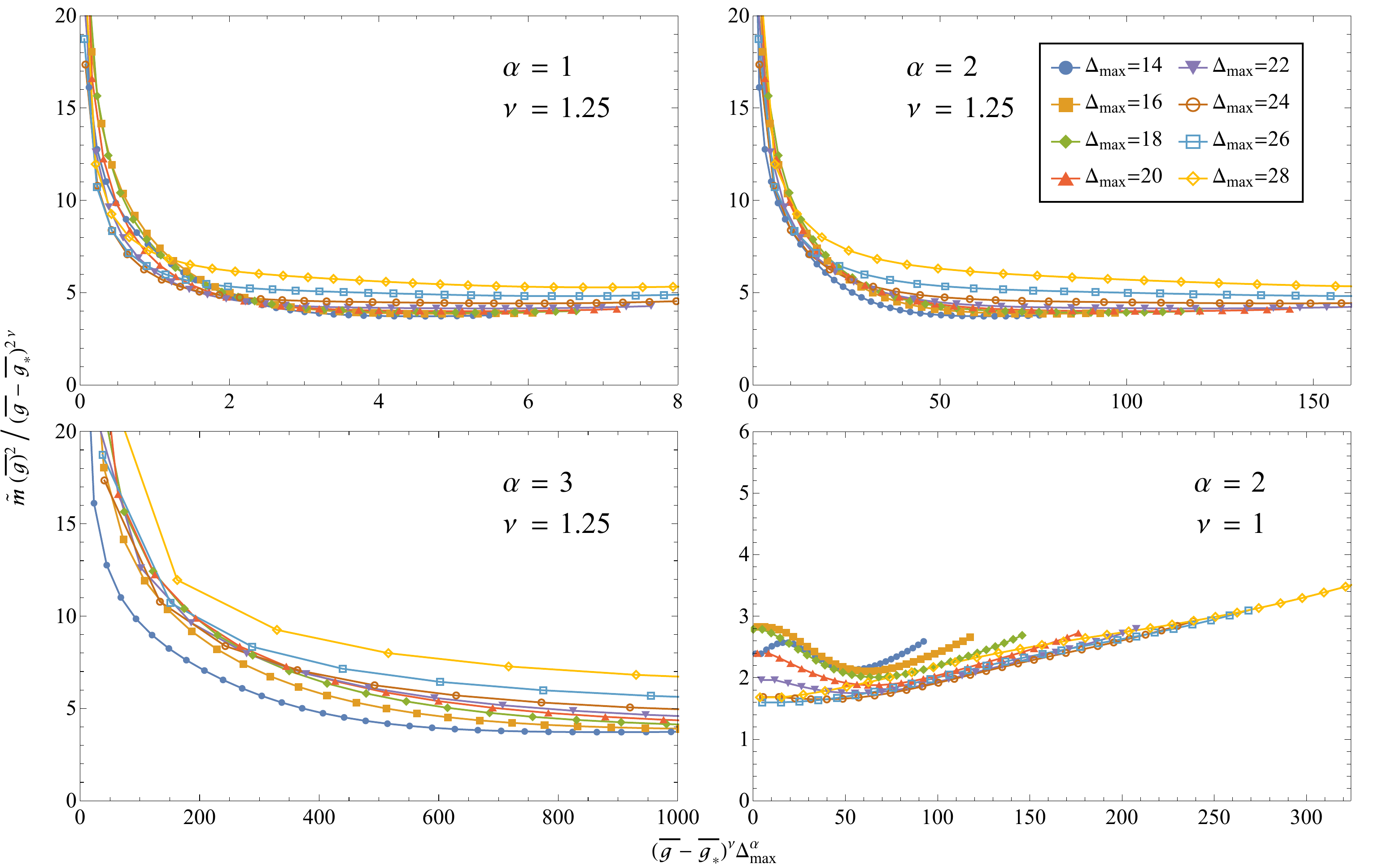}
\caption{
	\label{fig:universalIRMorePlots}
	Testing the universal IR scale model with different parameters.
	Fixing the critical exponent at the theoretical value, we plot the function with parameter $\alpha = 1$ (upper left), $\alpha=2$ (upper right) and $\alpha=3$ (lower left). Compared with $\alpha=2$, $\alpha=1$ has moderately less universal behavior across different $\dmax$ in the IR region (small $x$-axis), while $\alpha=3$ clearly does not have universality.
	Fixing the parameter $\alpha=2$ we also consider a different critical exponent, $\nu=1$ (lower right). Compared with $\nu=1.25$, there is no universal IR behavior in the small $x$-axis region and the ratio is also not constant above the IR scale (large $x$-axis).
}
}
 \end{figure}

In Section \ref{sec:universal-IR} we briefly mentioned that the universal IR scale model prefers the parameter $\alpha=2$ and $\nu=1.25$. In this appendix we provide the details by contrasting the prefered parameters with more general choices and argueing how the analysis discriminates them. Recall that in our model the IR behavior depends on a single scale 
\begin{align}\tag{\ref{eq:IR-scale}}
	\frac{\mgap}{\Lambda_{\rm IR}} \propto \dmax^\alpha (\bar g_*- \bar g)^{\nu} \, .
\end{align}
and we measure the behavior of the quantity $\frac{\tmgap}{(\bar g_* - \bar g)^{\nu}}$ which has the meaning of the ratio of the observed (truncation modified) mass gap to the true mass gap. We expect the relationship to be  
\begin{itemize}
\item When the mass gap is above the IR scale ($\dmax^\alpha (\bar g_*- \bar g)^{\nu} $ large), and the ratio is constant. 
\item When the mass gap is below the IR scale ($\dmax^\alpha (\bar g_*- \bar g)^{\nu} $ small), the mass gap deviates from the critical exponent and behaves uniformly as a function of the variable (\ref{eq:IR-scale}) across different $\dmax$.
\end{itemize}

In Fig.\ \ref{fig:universalIRMorePlots} we try different combinations of $\alpha$ and $\nu$ in the Universal IR scale model. 
We try two different values above and below the chosen parameter $\alpha=2$, and the plots show the IR universality is not as good as $\alpha=2$.
We also find that the model weakly prefers $\nu=1.25$. In particular we would like to contrast it with $\nu=1$. The plot of $\nu=1$ fails to realize both features. First, the ratio is not a constant above the IR scale, suggesting $\nu=1$ does not fit the critical exponent. Second, below the IR scale, curves of different $\dmax$ span out, demonstrating no universal behavior.
We emphasize that the preference on $\nu=1.25$ is weak and should not be used to determine the critical exponent.

\twocolumngrid

\bibliographystyle{apsrev4-1-nospace}
\bibliography{GNYBib}

\end{document}